\documentclass[a4paper, 11pt]{book}

\usepackage[top=5.0in]{geometry}
\usepackage{fullpage}
\usepackage{algorithm}
\usepackage{algorithmic}
\usepackage{listings}
\usepackage{graphicx}
\usepackage{float}
\usepackage{amsthm}
\usepackage{amssymb}
\usepackage{amsmath}
\usepackage{textcomp}

\usepackage{lscape}
\usepackage{multicol}

\usepackage{epigraph}
\usepackage{titlesec}
\titlespacing*{\chapter}{0pt}{-50pt}{20pt}
\titleformat{\chapter}[display]{\normalfont\huge\bfseries}{\chaptertitlename\ \thechapter}{20pt}{\Huge}

\usepackage{url}
\usepackage{verbatim} 
\usepackage{subfig}
\usepackage{wrapfig}

\usepackage{caption}
\usepackage[table]{xcolor}
\definecolor{tableShade}{HTML}{F1F5FA}
\definecolor{tableShade2}{HTML}{ECF3FE}
\definecolor{bg}{rgb}{0.95,0.95,0.95}

\usepackage{fancyhdr}
\newcommand{\centertitle}[1] {\begin{center} \Large \textbf{#1} \end{center}}

\usepackage{makeidx}
\makeindex

\usepackage{xcolor}
\usepackage{setspace}

\definecolor{shadecolor}{rgb}{1,0.8,0.3}

\newsavebox{\boxcontainer}

\newcommand{\mybox}[1]{
	\colorbox{gray!20}{
		\small
		\begin{minipage}[t]{\linewidth}
		\begin{spacing}{1.0}
		#1
		\end{spacing}
		\end{minipage}
	}
}

\usepackage{graphics}
\usepackage{epsfig}

\usepackage[bookmarksnumbered,pdfpagelabels=true,plainpages=false,colorlinks=true,
            linkcolor=black,citecolor=blue,urlcolor=blue,breaklinks=true,]{hyperref}
\newenvironment{abstract} {
  \newpage
  \thispagestyle{empty}
  \begin{center}
    \Huge
    Abstract
    \normalsize
  \end{center}
  }
{}

\newenvironment{acknowledgements} {
  \newpage
  \thispagestyle{empty}
  \begin{center}
    \Huge
    Acknowledgements
    \normalsize
  \end{center}
}
{}

\begin{document}
\frontmatter

\begin{titlepage}
\begin{center}
\includegraphics[width=200pt]{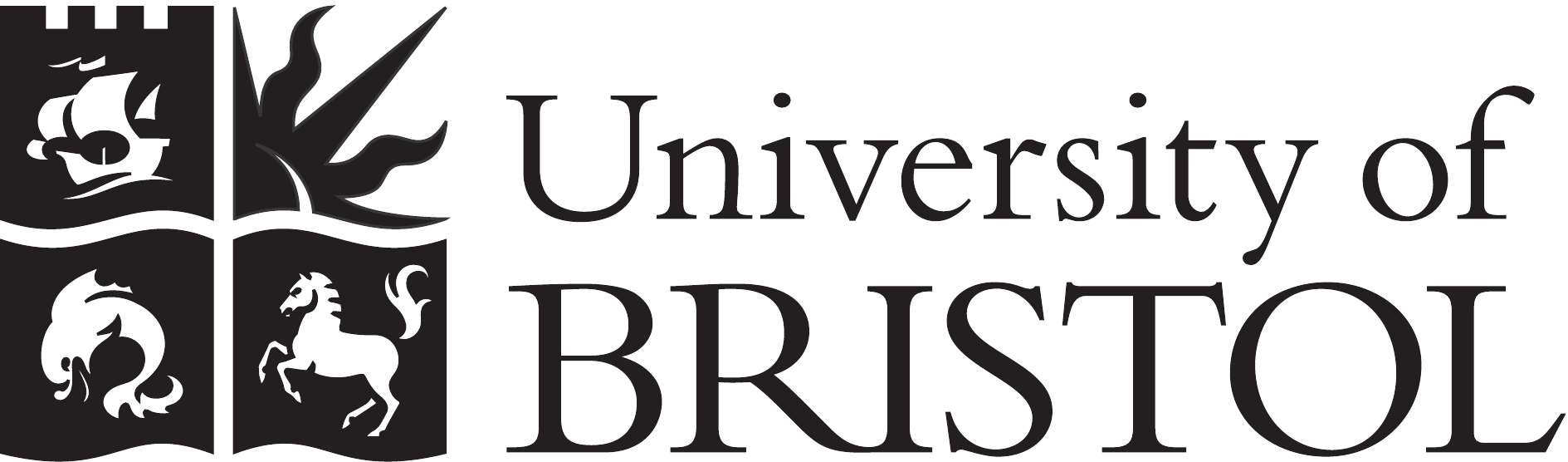}
\end{center}

\vspace{130pt}
\begin{center}
\Huge
Real-time jam-session support system.

\large
\textbf{Name:} Panagiotis Tigkas pt0326
\\
\textbf{Supervisor:} Tijl De Bie
\end{center}

\vspace{200pt}
\begin{center}
September 2011
\end{center}

\end{titlepage}
\let\cleardoublepage\clearpage


\doublespacing

\begin{abstract}
We propose a method for the problem of real time chord accompaniment of improvised music. Our implementation can learn an underlying structure of the musical performance and predict next chord. The system uses Hidden Markov Model to find the most probable chord sequence for the played melody and then a Variable Order Markov Model is used to a) learn the structure (if any) and b) predict next chord. We implemented our system in Java and MAX/Msp and compared and evaluated using objective (prediction accuracy) and subjective (questionnaire) evaluation methods. Our results shows that our system outperforms BayesianBand in prediction accuracy and some times, it sounds significantly better.
\\
\\
\textbf{keywords}: Machine Learning, Interactive Music System, HCI
\end{abstract}

\begin{acknowledgements}

I would like express my deepest gratitude and thank my supervisor, Tijl De Bie, where without his guidance and his comments I wouldn't be able to complete this project.

Most importantly, I would like to thank my parents for being my sponsors and supporters of my decisions and the fact that without their love and help I wouldn't be able to fulfil my dreams.

\end{acknowledgements}

\newpage
\thispagestyle{empty}
\begin{center}
 \Huge
 Declaration
 \normalsize
\end{center}
This dissertation is submitted to the University of Bristol in accordance with the requirements of the degree of Master of Science in the Faculty of Engineering. It has not been submitted to any other degree or diploma of any examining body. Except where specifically acknowledged, it is all the work of the Author.
\\
\\
\\
\noindent
Panagiotis Tigkas\\
September 2011

\newpage

{\fontsize{3.4mm}{3.4mm}\selectfont \tableofcontents}

\mainmatter

\onehalfspacing


\chapter{Introduction}

\epigraph{\emph{Good afternoon, gentlemen. I am a HAL 9000 computer. I became operational at the H.A.L. plant in Urbana, Illinois on the 12th of January 1992. My instructor was Mr. Langley, and he taught me to sing a song. If you'd like to hear it I can sing it for you.}}{2001: A Space Odyssey}

\section{Motivation}

My studies on machine learning were driven by the question of whether machines are capable of learning like humans,
acting intelligently or interacting with humans for the accomplishment of a task. Furthermore, as a musician I was challenged by the idea of whether machines are capable for creativity, either as autonomous agents or by interacting with human performers.

The idea of developing a system that join a jam session and support musicians by providing accompaniments 
or improvising music came from my need to explore and understand the way that humans improvise music. Trying to mimic
the way a young musician learn to play or improvise music, we researched and developed a system which is capable of providing chords to an improvising musician in a jam session in real-time.

In this thesis, we utilised supervised machine-learning methods which have been successfully used in fields like
computational biology, text mining, text compression, music information retrieval and others. The process we approached this problem can be summarised in the following sentence. Interacting with improvising musicians in real-time using experience learned from data (off-line) and the rehearsal (on-line).

\section{Goals and contributions}

The main goal of this thesis is the development of a system that will be able to infer an underlying structure of the improvisation
and predict and play chord accompaniments. A challenge of such system is that it must work under the real-time constraint ; that is, it must 
predict next chord and play it without the musicians (or the audience) noticing artificial latencies. One other issue that makes such system challenging is that the input of the system is melody. This introduce further complexity to the problem since there is no strict mapping from melody to chords. What is more, the melody on itself doesn't contain sufficient information to provide "correct" chords. Thus, we need a subsystem that will be able to extract from a melody an underlying structure. That is, a chord progression that best explains/match the melody. Such system, however, might introduce errors that get propagated to the predictor, thus careful design of both subsystems (inferencer and predictor) is needed.

Analytically, the main objectives of this project are:
\begin{itemize}
\item To develop a Hidden Markov Model using Viterbi algorithm that given a melody, will be able to infer the corresponding chords (from time $0$ to $t$).
\item To use the information from the Hidden Markov Model and a Variable Order Markov Model to predict next chord (time $t+1$).
\item To develop current state-of-art system which is based on Bayesian Networks \cite{Kitahara2009} so as to compare.
\item To evaluate the system using both objective and subjective evaluation.
\end{itemize}

\begin{figure}[h!tp]
\centering
\includegraphics[width=\linewidth]{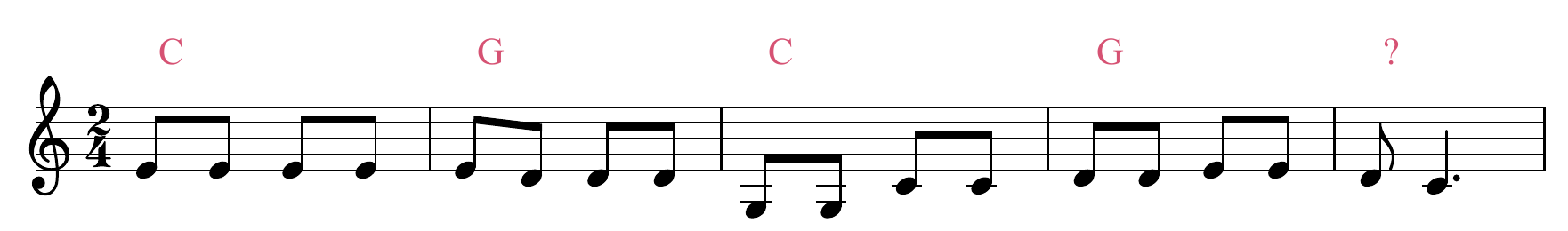}
\caption{Chord prediction given a melody task. The chords in red are the chords as found from Hidden Makrov Model. The question mark indicates that we have to predict that chord}
\end{figure}

Our contribution with this thesis is the development of such system which is capable of both off-line and on-line learning, like a musician
which train himself with practice songs (off-line learning) and also understand the structure and the tensions in a jam-session
and adapt the performance (on-line learning). As a product of this thesis, a plugin and a standalone application was 
developed as a Java external in Max/MSP and Ableton live \footnote{Max/MSP is a visual programming language for multimedia and music programming. Ableton live is software for real-time music performance and composition.}. What is more, as byproduct
of the thesis we developed a framework for creating online questionnaires, parsers for MIDI and MusicXML files and
several python scripts for statistical processing of musical data. A complete list and the repository of the files is given in the appendix of the thesis.

\section{Outline}
In the following chapter we will introduce the reader to the field of interactive music systems and the methods which we used during this thesis. What is more, with respect to the non-musician reader, we provide a musical background which will be sufficient for the understanding of the thesis.

In chapter 3 we describe our system and present the design choices we made so as to accomplish our project. What is more we present the settings and the topology of the used models and give a description of the data set used for training and testing of our system.

In chapter 4 we present the results of the evaluation of our system. 

Finally in chapter 5, we give an interpretation of the results, discuss the project's contribution and present a plan for future work.
\chapter{Background}

\epigraph{\emph{The old distinctions among emotion, reason, and aesthetics are like the earth, air, and fire of an ancient alchemy. We will need much better concepts than these for a working psychic chemistry.}}{Marvin Minsky}

In this chapter we aim to provide a brief brief overview on related work and state-of-the-art in interactive music systems and we give an introduction to the reader on music theory and computer music.

\section{Interactive Music Systems}

Robert Rower coined the term Interactive Music Systems to describe the upcoming subfield of Human-Computer Interaction
where machines and humans interact with each other in musical discourse. Music is the product of such interaction where 
the computer takes the role of either the instrument (e.g. wekinator) or the fellow musician (solo improvisation, chord supporter, etc).

Unfortunately, taxonomy of such systems is still under development since there are several disambiguations with terms such as "interaction" or "musical systems". As Drummond \cite{Drummond2009} coins, there are cases that the term "interactive system" describes reactive systems that contain the participation of audience, where the main difference between reactive and interactive is the predictability of the result. In this section we aim to give a brief description of related work in interactive musical systems using a simplification of Rowe's taxonomy.

According to Rowe \cite[p7-8]{rowe1992interactive} Interactive Music Systems can be categorised using the following three dimensions:
\begin{enumerate}
\item \textbf{score-driven or performance-driven:} The performance which the system interact with is precomposed or impromptu (without preparation)
\item \textbf{Transformative, generative or sequenced}: The system either transforms the input or generates novel music or playback of stored material
\item \textbf{Instrument role or player role:}  The system is an extension of the human performance or
an autonomous entity in the performance.
\end{enumerate}

\begin{figure}[h!]
\centering
\includegraphics[width=0.8\linewidth]{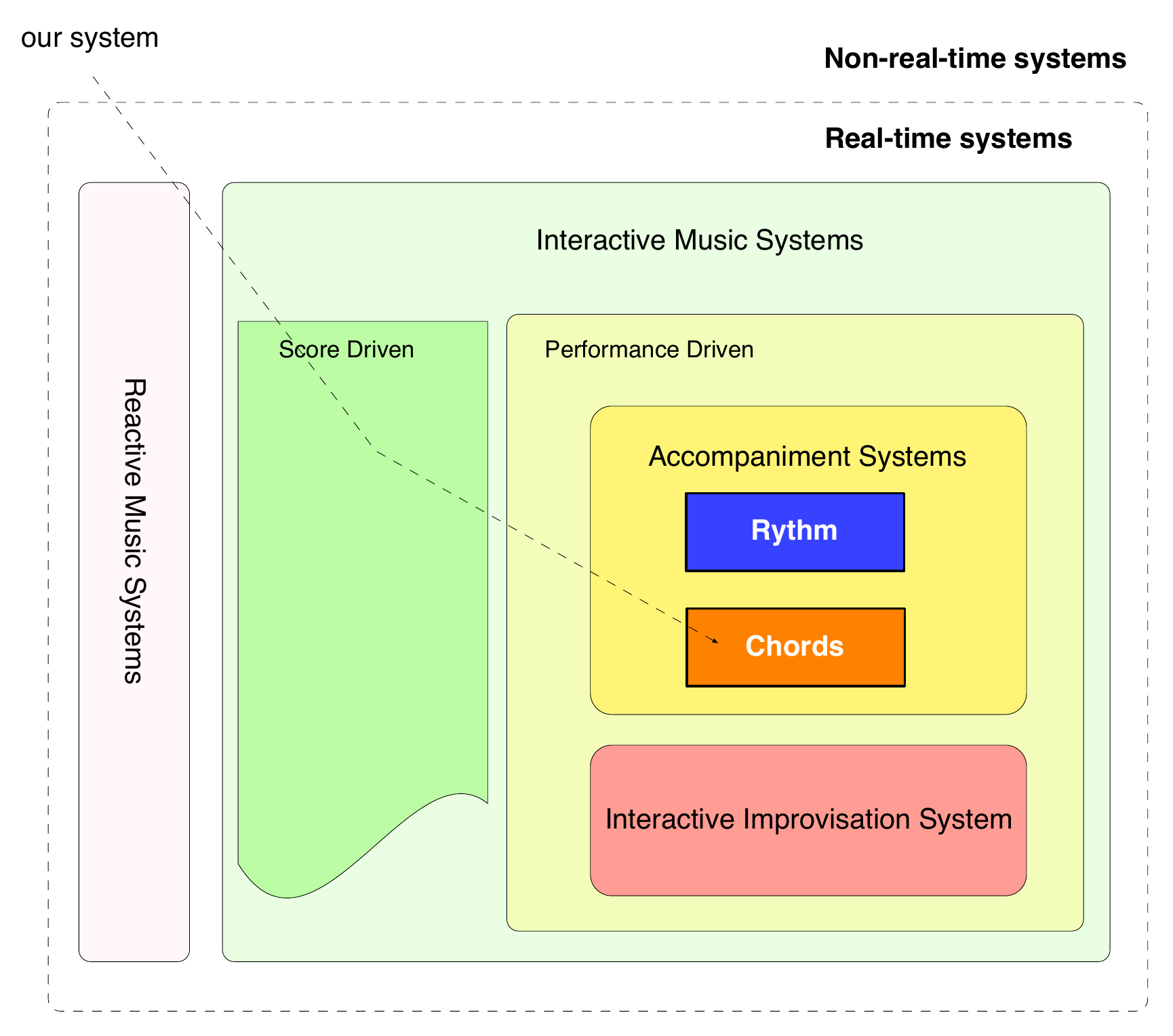}
\caption{Taxonomy}
\label{taxonomy}
\end{figure}

Covering the cartesian product of those dimensions is out of the scope of this thesis. However, we will use the score-driven vs performance-driven
dimension into our taxonomy (figure \ref{taxonomy}).

\subsection{Score Driven}

\textbf{Music Plus One} \cite{Raphael2010}

Christopher Raphael \cite{Raphael2010} developed a real-time musical accompaniment system for oboe.
The system interacts with a musician who plays a \textbf{non-improvised} music and provides accompaniments
using precomposed information. For this, the system use Hidden Markov Model so as to follow a score at real-time.

In detail, Music Plus One consists of three subsystems called "Listen", "Predict" and "Play". The first subsystem
take as input the audio of the musician and analyse it so as to find onsets. However, "Listen" subsystem might introduce latencies that are crucial for the interactivity of the system. Thus, "Predict" subsystem tries to predict next events and thus improve system's response. Prediction gets improved as long as new information becomes available from "Listen" subsystem. "Predict" subsystem make use Gaussian mixtures. Finally, "Play" subsystem output audio by using phase-vocoding technique \cite{flanagan1966phase}.

\subsection{Performance Driven}

\subsubsection{Interactive Improvisation System}

\textbf{Imrovised Music with Swarms} \cite{Blackwell2002}

Blackwell \cite{Blackwell2002} criticised previous systems for their lack of interaction with the musicians. What is more, he believes that they were mainly focused in modelling and encoding musical knowledge, providing to the user accompaniments or suggestions regarding harmony or compose algorithmically using the knowledge and rules (aesthetics) hardcoded from the programmer. His contribution was to develop a system using ideas inspired by the way bird organised in flocks.

Mimicking swarms' for creating intelligence is not a new idea (similar idea to Minsky's Society of Mind), however it is the first time that it is applied for interaction and, moreover, in musical interaction. Music like flocks is a self-organizing system. Harmony, melody, chords are moving towards directions with respect an organisation. Musicians force the flock to different directions without breaking this organisation. Blackwell, in this work, created a system using original ideas initially developed by  Craig Reynolds in his program "Boids" \cite{Reynolds1987}.

\textbf{Genjam} \cite{Biles}

John Biles in GenJam utilise genetic algorithms for jazz improvisation. Although it's original version in 1994 was not interactive in 1998 the system re-designed so as to improvise in a real-time context.

Genetic Algorithms originally developed by Holland in 1975 \cite{holland1975adaptation} are biological inspired algorithms which mimic the evolution process in nature. They consist of a set of encoded strings called population. In GA terminology the strings are called genotypes. The algorithm evolve the population by applying mutations and combinations of the genotypes iteratively. Also, each time it evaluates the population by using an objective function, called fitness function.

The system performs solo trading, which means that it respond to a melodic phrase by mutating it. The approach followed in GenJam is called "fours trading". That is, it listens the last four measures from the human participant and it saves it in a chromosome. Then, it calls a Genetic Algorithm which after several evolutions it produces a new population / music solo.

\textbf{A real-time genetic algorithm in human-robot musical improvisation} \cite{Weinberg2009}

Weinberg \emph{et al.}, developed a system in 2009 which is able to improvise melodies in an interaction with a human musician. Their system actually
is a robotic arm which plays a xylophone. This system also make use of genetic algorithms so as to to improvise. 

The system is fed with melodies which then fragments and uses as population. Similarly to GenJam, each time the system is called to improvise
it mutates and combines several genes from the population and evaluates them in means of how much they "fit" the melody which the system is called
to answer. The method they used to calculate fitness is Dynamic Time Warping \footnote{Dynamic Time Warping is a technique for measuring similarity in patterns that vary in time.}, which is very similar to Levenshtein distance \footnote{Levenshtein distance is metric used in calculating strings similarity.}.

\textbf{ Continuator \cite{Pachet2003} }

Continuator was one of the greatest achievements in the filed of Interactive Music Systems and inspired this thesis in great means. Pachet's system performs in an improvisation and is capable of providing continuations of a soloing melody in the same style. This is of great importance since it's the first time that style agnostic system achieves to retrieve and simulate the participants style. What is more, his work showed the ability of variable order Markov models to work in real-time context. We will describe the model and the system in detail in next chapter where we introduce variable order Markov models.

\subsubsection{Accompaniment systems}

\textbf{ BayesianBand \cite{Kitahara2009} }

BayesianBand is a system which was developed by Kitahara et al in 2009. Their system provides accompaniments in improvised music using Bayesian networks in real time. As far as we know, this is the closest system to ours and thus we used it to compare with.

In their approach they use Bayesian networks so as to infer next note and then predict next chord. What is of great importance is that they update the probabilities  during performance so as to improve next note inference. That way they improve also prediction of next chord since, as we will se later, next chord depends on previous notes and chords predicted. We won't describe it in detail until we describe Bayesian networks, in chapter 3. 

\section{Musical background}

\subsection{Elements of music theory}

\subsubsection{Notes}
A musical composition is very similar to a construction. Notes are the building blocks of (almost) every musical composition. They are the "Atoms" of Western music.

Musical notes consist of 5 properties. Pitch, note value, loudness, spatialization and timbre. We will mainly focus in pitch and value. Briefly, loudness is how loud a note is perceived, spatialization is the position in the space (e.g. left/right) and timbre is the quality of the sound, the texture.

\begin{figure}[h!]
\centering
\rowcolors{1}{white}{tableShade}
\begin{tabular}{l|l}

Pitch class & Frequency (Hz) \\
\hline
A & 440.00 \\
A\# &  466.16 \\ 
B & 493.88 \\
C & 523.25 \\
C\# & 554.37 \\
D & 587.33 \\
D\# & 622.25 \\
E & 659.26 \\
F & 698.46 \\
F\# & 739.99 \\ 
G & 783.99 \\
G\# & 830.61 \\ 

\end{tabular}
\caption{Note to frequency }
\label{note2freq}
\vspace{-10pt}
\end{figure}

Pitch is how human perceives sound frequencies. For example a sound of 440 Hz in Western music is perceived and assigned to the pitch A. Usually we will refer to the pitch as pitch class since each pitch is a class of frequencies and not a specific frequency. More specific, human ear perceives as in the same same frequencies whose ratio $2^n, n\in\mathbb{Z}$. For example a sound with $f=440 Hz$ is perceived in the same pitch-class as a sound with $f=880 Hz$. In figure \ref{note2freq} \footnote{\url{http://www.cs.nyu.edu/courses/fall03/V22.0201-003/notes.htm}} you can see a mapping of a frequency to a pitch class. More formally, a pitch class of a frequency A is:
\begin{align}
\text{PitchClass}(A) = \{ B: A/B = 2^n, n\in\mathbb{Z} \}
\end{align}
However, we will refer to pitch class by a symbolic name, e.g. A is the pitch class which contains the  $2^n, n\in\mathbb{Z}$ multiplications of 440 Hz frequency.

We won't talk about frequencies anymore. From now on we will only concern about pitch classes notated after their english name ( $\{[A-G]\}\cup\{\flat,\sharp,\_\}$ ).

The difference between two notes is called interval. The smallest interval is a \emph{semitone} and two semitones make a \emph{tone}. In Western music, a semitone is $\frac{1}{12}$ of an octave, were octave is the minimum distance between two different pitches which are in the same pitch class. For example a note with a pitch of frequency 880 Hz is an octave higher than a note of frequency 440 Hz.

Another element of Western music are the \emph{accidentals}. The flat accidental $\flat$ decrease the note which is applied to by one semitone and $\sharp$ increase the note by one semitone. For example $A\sharp$ is pitch class $A$ raised by one semitone. As you can see, in figure \ref{note2freq} we included accidental notation to the pitch class. 

Also an assumption we make which we will discuss later is that two letters without accidentals have a distance of a tone, except E-F and B-C which have a distance of a semitone. That is because as we will see, there is a specific arrangement of the notes in frequency space. 

\begin{figure}[h!]
\centering
\includegraphics[width=0.6\linewidth]{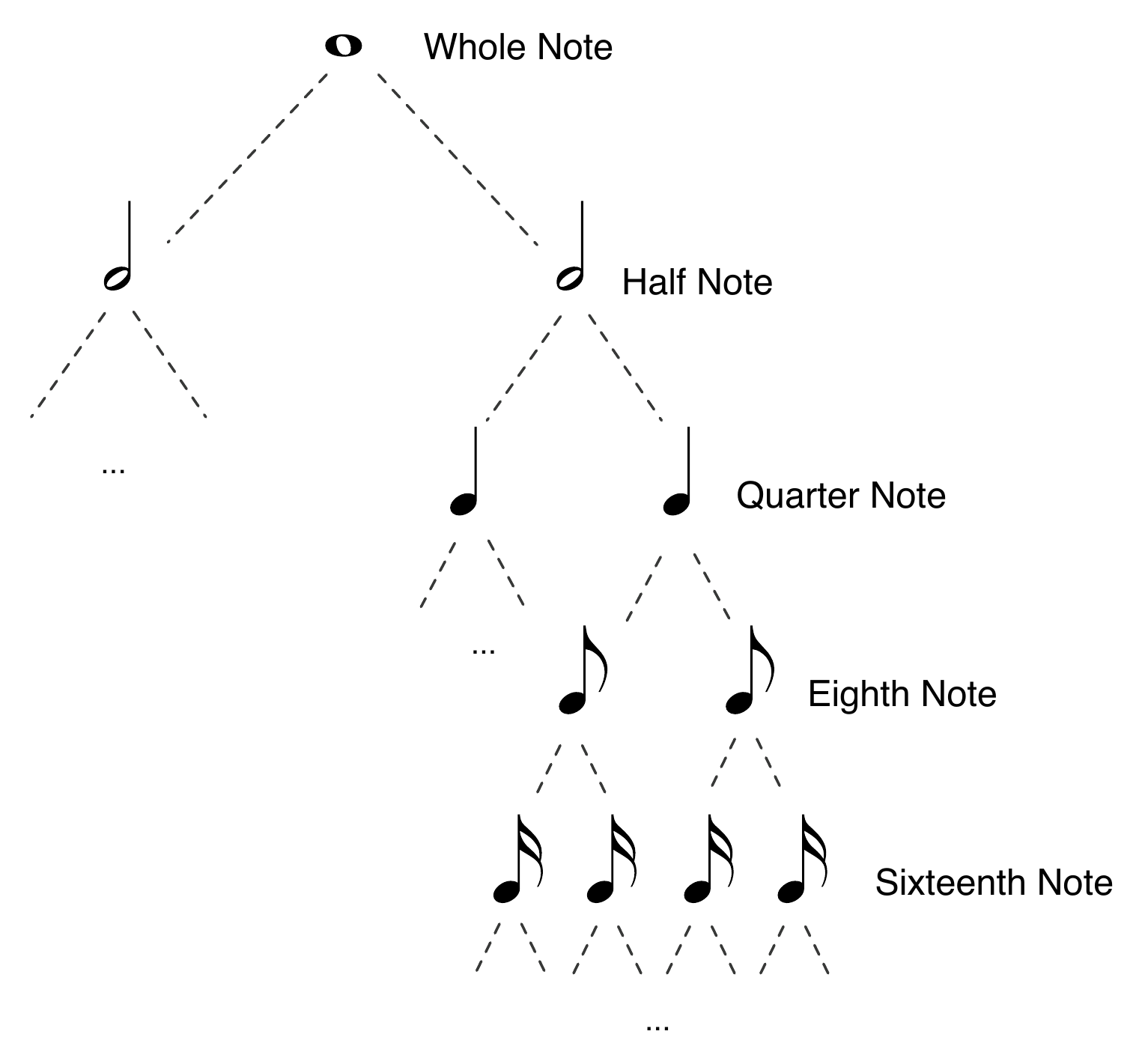}
\caption{Note hierarchy}
\label{nothi}
\end{figure}

Another characteristic of a note we will discuss is \emph{note value}, which is the duration in time or it's relative time distance from the next note. A whole note is a note which length equals to four beats. As we go deeper in the tree in figure \ref{nothi}, the duration is subdivided. For example a Half Note lasts two beats,
a Quarter one beat, etc.

Finally another very important element of music is silence which only characteristic is duration. The musical name of silence is \emph{rest} and it's value has the same meaning with notes' duration. Note value is what gives rhythmic meaning in a musical composition. Although there are a lot more to be said about notes we will stop here since further explanations are out of the scope of this thesis.

\subsubsection{Rythm}

Notes are arranged in frequency space but also in time. This arrangement in time is what humans perceive as rhythm. A musical composition is characterised by a \emph{tempo}. It's the heartbeat of music and is measured in \emph{Beats Per Minute} (BPM). Notes are grouped in \emph{meters} (also called \emph{bars}). Each meter contain notes which duration sum is the same in each meter. In other words, meters split the composition in equally sized groups. The size of a group is in means of notes duration.

Time signature specifies how many beats are in each meter. For example a common time signature is $\frac{4}{4}$ which means that we have \textbf{four quarters of note}, where unless differently defined, a quarter is one beat.

\vspace{-14pt}
\subsubsection{Melody and Chords}
Melody is a sequence of notes of several pitches and time values. In Western music melody usually consists
of several phrases or motifs or patterns.

A chord is group of two or more notes that are played simultaneously. A typical form of chords consists
of three notes (triads).  A triad usually consists of the root, the third and the fifth, where the root is the first note of the chord. The third is the note which distance is three (minor) or four (major) semitones from the root and a fifth is a note which distance is 7 semitones. The quality of the third separates minor chords from major chords. It's easy to see that for 12 notes we have $12+12=24$ triads (minors and majors).

In a musical composition, chords are usually played in progressions revealing structures and patterns. Chord progressions and melody are highly dependent and each restricts the other under the rules of harmony. 
\vspace{-14pt}
\subsubsection{Tonal music, scales and Harmony}

Tonal music is the music which is organised around chord progressions. However, this doesn't make it special by it self. We need also a constraint of what chords are allowed and what not. This constraint is the scale. 

A scale, in simple words, is a set of notes that we are allowed to play. Under this constraint melody and chords are developed.   In an improvisation act, musicians share and respect this constraint. It is the common ground. Scale usually characterise a composition, however, there are cases where the scale change (key modulation) and thus new constraints are introduced.
\vspace{-14pt}
\subsection{Computer music}

Finally, we will close this introduction to the reader with a brief description of some elements of computer music.
\vspace{-14pt}
\subsection{MIDI protocol}

So far we saw how music is decomposed. However, we need a way to model music so as to be understandable by computers. Please note that by music we mean symbolic music, the blueprints of a musical composition and not the audio.

Musical Instrument Digital Interface (MIDI) is a protocol which describes music in sequences of events. Those events can be streamed from an instrument (e.g. synthesiser) or saved in files. The most important events are those
of \textbf{Note On} and \textbf{Note Off}. Each of those events have the specific format.

\[
\text{MIDI event: } \underbrace{\phantom{[}t\phantom{[}}_{\text{Time elapsed}}, \underbrace{\phantom{[}0\text{x}80	\text{ or } 0\text{x}90\phantom{[}}_{\text{MIDI event type}}, \underbrace{[0-127]}_{\text{Note Number }}, \underbrace{[0-127]}_{\text{Note Velocity }}
\]

As we can see, midi events are sufficient to describe pitch classes, durations and time distances between notes. Chords can be seen as sequences of notes with $t=0$.

\begin{figure}[h!]
\centering
\begin{minipage}{70pt}
\begin{verbatim*}
...
1200,note on,41,90
1392,note off,41,0
1440,note on,41,90
1512,note off,41,0
1530,note on,41,90
...
\end{verbatim*}
\end{minipage}
\caption{example of midi events}
\end{figure}

\vspace{-14pt}
\subsubsection{Programming languages and tools}
So as to simplify the process in our thesis and focus in machine learning algorithms and methods we aim to make use of a audio/music programming language. After researching we found that the most appropriate language for this task was Max/MSP which is a visual programming language, originally developed by  Miller Puckette in IRCAM and commercially developed by Cycling'74 \footnote{ \url{http://cycling74.com/} }.

Our methods will be implemented mainly in Java language since we can easily develop plugins for the mentioned programming language.

Finally, for music synthesis we aim to use Ableton Live \footnote{\url{http://www.ableton.com}} which is a suite heavily used in computer music creation.

\chapter{Graphical Models}

A graphical model, is a statistical model for describing dependencies between random variables which are expressed in terms of conditional probabilities. Conditional probability represents the probability of an event given the knowledge of another event. For example 
\begin{align}
P(\text{"chord is Em"} | \text{"last note played is E"})
\end{align}
denotes the probability that the chord to be played is E minor given that the last note played was E.

As the name reveals, a graphical model is a graph $G=(V,E)$, whose nodes $V$ are the random variables and the nodes $E$ describe the dependencies among nodes $V$. \cite[ch. 3]{pearl1988probabilistic} and \cite[p.359]{bishop2006pattern}.

The advantage of graphical models is that they simplify and reduce the complexity of the computation of joint probabilities in means that we can easily compute them by decomposing them into a product of factors. We will now see a specific class of graphical models named Bayesian networks.

\section{Bayesian networks}

\begin{figure}[h!]
\centering
\includegraphics[width=180pt]{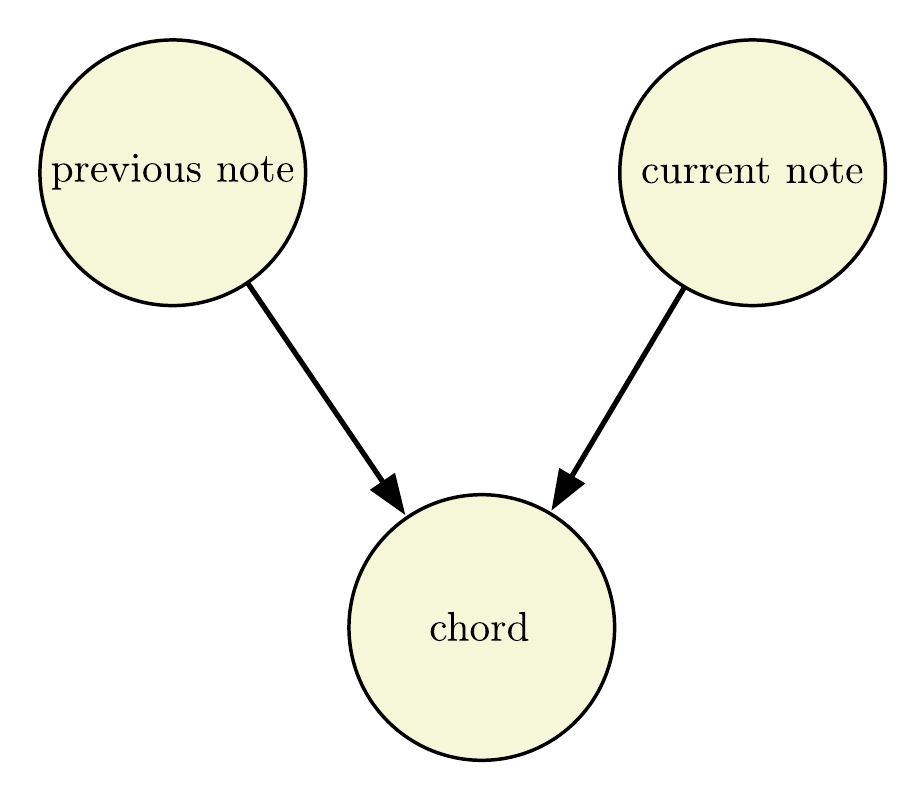}
\caption{Example of graphical model}
\label{graphicalmodel}
\end{figure}

Bayesian networks are graphical models with the extra property that the graph is directed and acyclic. Each node represents a random variable in means of Bayesian statistics. The edge of the graph represents dependency of the random variables and the direction show which variable depends on which. For example, in figure \ref{graphicalmodel}, the model is interpreted as following. The probability of a chord playing is conditioned from the probability of a note currently playing and the note played before.

Bayesian networks are very important due to the simplification that they introduce in computing join probabilities. One should sum over all possible different values of  random variables . For example for a graphical model with binary random variables $A,B,C,D$ and dependency chain $A\rightarrow B\rightarrow C\rightarrow D$, the joint probability $P(A,B,C,D)$ is computed as \[ P(A,B,C,D)= P(A)P(B|A)P(C|A,B)P(D|A,B,C) \]
Without using the topology of the dependency graph, this computation might be inefficient. Bayesian networks provide a framework for simplifying the computations exploiting the dependencies.

Let $pa(n)$ be the function which returns the parents of a node $n$. For example in figure \ref{graphicalmodel}, $pa(\text{chord})$ is $\{\text{current note}, \text{previous note}\}$ (random variables of the graph).

Then for $\text{x}=\{x_1, x_2,\dots,x_K\}$ the joint probability $Pr(\text{x})$ is computed as:

\begin{align}
Pr(\text{x})=\prod_{i=1}^{K}{Pr(x_i | pa(x_i))}
\end{align}

For examples, for the model in figure \ref{graphicalmodel} we have:
\begin{align} \label{eq:example}
& Pr(\text{Chord}, \text{Current note},\text{Previous note})=\\
& Pr(\text{Chord} | \text{Current note},\text{Previous note})Pr(\text{Current note})Pr(\text{Previous note})
\end{align}

When the random variables represents sequence of data (e.g. time series or sequence of data like proteins) then the network is called \textbf{Dynamic Bayesian Network}.

\subsection{Applying bayesian networks for chord prediction}

In the previous section, we saw how a graphical model can capture the dependencies among random variables with graphical models and we introduced a specific class of graphical models, the \emph{Bayesian networks}.
As we will see, we can use Bayesian networks to provide a fast solution for the problem of providing supporting accompaniments to an improvising melody. The method
we describe is based on the BayesianBand system as proposed by Kitahara in 2009 \cite{Kitahara2009}.

A musician during a jam-session, tries to predict the music which will played in the future, based on his experience during the session. For example, a guitar player supports a fellow musician by playing chords which will be consistent with the melody. To do so, the player will try to predict the note which will be played and then decide which chord is more appropriate. This process can be decomposed into two phases:
\begin{itemize}
\item \emph{Melody prediction phase} where the musician  predict the note which will be played next. 
\item \emph{Chord prediction / inference phase} where the musician decide which chord fit on the melody playing and the predicted note.
\end{itemize}


\subsubsection{Inference using bayesian network}

\begin{figure}
\centering
\includegraphics[width=170pt]{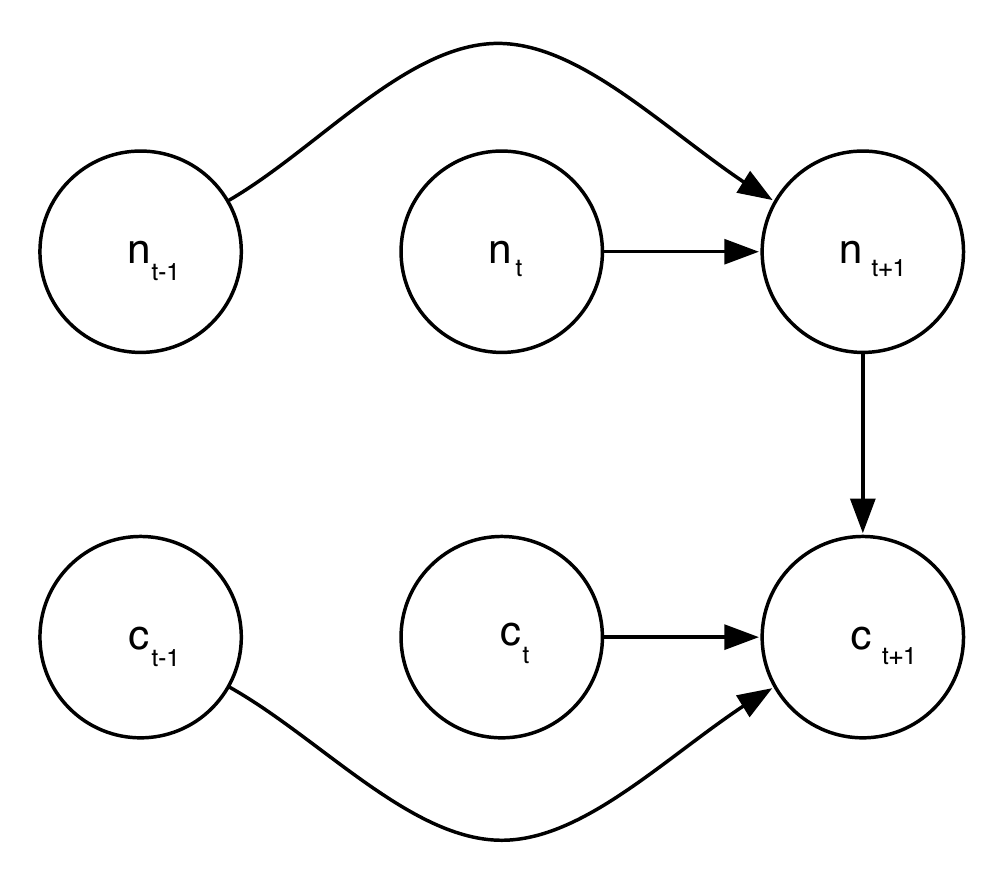}
\caption{Bayesian network}
\label{bayesianband}
\end{figure}

Lets define a chord sequence $\textbf{c}$ as
\[
\textbf{c} = (c_1,\dots,c_n)
\]
and a melody $\textbf{n}$ as a sequence of notes
\[
\textbf{n}=(n_1,\dots,n_m)
\]


As we saw in the music theory section, there is a sequential dependence in melody and chord progressions. Thus
\begin{align}\label{eq:notes}
Pr(n_{t+1}|\textbf{n})
\end{align}
 and 
\begin{align} \label{eq:chords}
 Pr(c_{t+1}|\textbf{c})
\end{align}
 
So as to simplify the process, Kitahara \emph{et al.} \cite{Kitahara2009} assume that notes and chords are 2nd order Markov chains. In that way, the complexity of the algorithm is kept low and thus a system which will decide and predict in real time can be developed. 

Under the Markov assumption probabilites \ref{eq:notes} and \ref{eq:chords} are simplified to
\begin{align} \label{eq:notes2}
Pr(n_{t+1}|n_{t},n_{t-1})
\end{align}
 and 
\begin{align} \label{eq:chords2}
 Pr(c_{t+1}|c_{t},c_{t-1})
\end{align}


The Bayesian model shown in figure \ref{bayesianband} is driven by the following observations. First of all there is a sequential dependence in the melody 
and the chords progression. What is more, a chord depends on the note which it supports, hence, we add an edge from predicted note ($n_{t+1}$) to next chord ($c_{t+1}$).

Recalling the task, the system must:

1) Predict the note that maximise the probability $Pr(n_{t+1}|n_{t},n_{t-1})$
\begin{align}
o=\arg_{o}{\max{Pr(n_{t+1}=o|n_{t},n_{t-1})}}
\end{align}

2) Then, find the value of $c_t$ latent random variable which maximise the probability
\begin{align}\label{eq:myprob}
c=\arg_{c}{\max{Pr(c_{t+1}|n_{t+1}=o,c_{t}=co_1,c_{t-1}=co_2)}}
\end{align}
where $co_1$ and $co_2$ are the values as observed.

Please note each time there is a new note, the predicted chord is played and then predict  next note. That way
 time-response of the system is reduced. What is more, each note gets 12 note values, one for each pitch-class. Also, the chords they use are the 7 diatonic chords (chord allowed under the key constraint).
 
\subsubsection{Incremental update}
A main property in a jamming session is novelty and interaction \cite{Pachet2003}. That means that during the session, musicians should learn what fellow musicians play. In this case, the melody prediction should use knowledge from the session and adapt the prediction on the current melody playing. So as to achieve that, they update their model each time a new note of the melody is played. 

Thus,  the probability $Pr(n_{t+1}|n_{t},n_{t-1})$ is computed as follows:

\begin{equation}
Pr(n_{t+1}|n_{t},n_{t-1}) = \frac{ 
Pr_{\text{corpus}}(n_{t+1}|n_{t},n_{t-1})+\alpha\log{N(n_{t-1},t_{t})} \frac{N(n_{t-1},t_{t},n_{t+1})}{N(n_{t-1},t_{t})}
}{
1+\alpha\log{N(n_{t-1},t_{t})}
}
\end{equation}

where $N(n_{t-1},t_{t})$ and $N(n_{t-1},t_{t},n_{t+1})$ are the frequencies that the notes $n_{t-1},t_{t}$ and $n_{t-1},t_{t},n_{t+1}$ appeared in that sequence during the session. Constant $\alpha$ is a parameter of the user and defines the novelty of the system.

Probability $Pr_{\text{corpus}}(n_{t+1}|n_{t},n_{t-1})$ can be easily computed using the corpus.

\subsection{Assumptions, limitations and extensions}

One of the greatest advantages of this system is that it can easily adapt to the performance. That comes from the fact that the model is characterised by the a-mnemonic property of Markov chains. However, that assumption make also impossible to
understand and follow a structure of the chord or note (melody) progressions. In our project we aim to improve this model so as to be able
to catch higher order dependencies. So as to accomplish that we aim to use Hidden Markov Model with variable order Markov chains.

\section{Markov Model}

Markov models are graphical models where each node represents the state of a random variable at a specific time and
each node depends only on the node corresponding to the immediate previous state. For example, we denote $X_t$ the random variable which
corresponds on a state at time $t$. The memoryless property (also called Markov property) states that the random variable corresponds at time $t$ depends only on the random variable at time $t-1$, or more formally:
\[
P(X_{t+1}|X_1,\dots|X_t)=P(X_{t+1}|X_t)
\]

\begin{figure}[h!]
\centering
\includegraphics[width=0.6\linewidth]{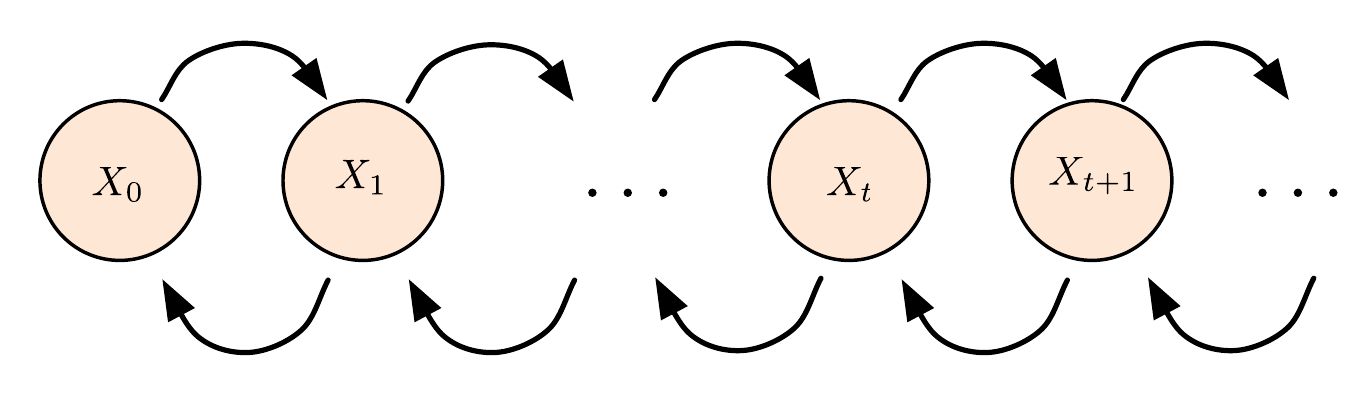}
\caption{Example of markov model}
\label{markov_model}
\end{figure}

\noindent
The possible values of $X_t$ is called state set $S$.

\noindent
Transition probabilities $P(X_{t+1}|X_t)$ are usually modelled using a transition matrix where $a_{ij}=P(X_{t+1}=S_i|X_t=S_j)$, where $S_i,S_j\in S$.

\noindent
One variation of Markov models is when each state depends on the previous $n$ states. More formally that means that 
\[
P(X_{t+1}|X_1,\dots|X_t)=P(X_{t+1}|X_t,X_{t-1},\dots,X_{t-n+1})
\]
This variation is called n-order Markov model. For the shake of simplicity 1st order Markov models are called simply Markov models.

\noindent
Until now we have seen some graphical models that will help us to describe the models that we used during this thesis.


\section{Hidden Markov Model}

Imagine that a Markov process (that is a stochastic process with the Markov property) generates data and we, as observers, observe not the actual process but the outcome of the process. This assumption is very important since we can simplify several tasks, as we will see later. Hidden Markov models are an extension of Markov models (sometimes known also as observable Markov models) with which we try to deal with such cases. 


Hidden Markov Models are a subclass of Dynamic Bayesian Networks and have been used successfully in speech recognition \cite{Rabiner1989}, in computational biology \cite{Krogh1994}, in prediction of financial time series \cite{Zhang2004} , etc

\begin{figure}[h!]
\centering
\includegraphics[width=1.1\linewidth]{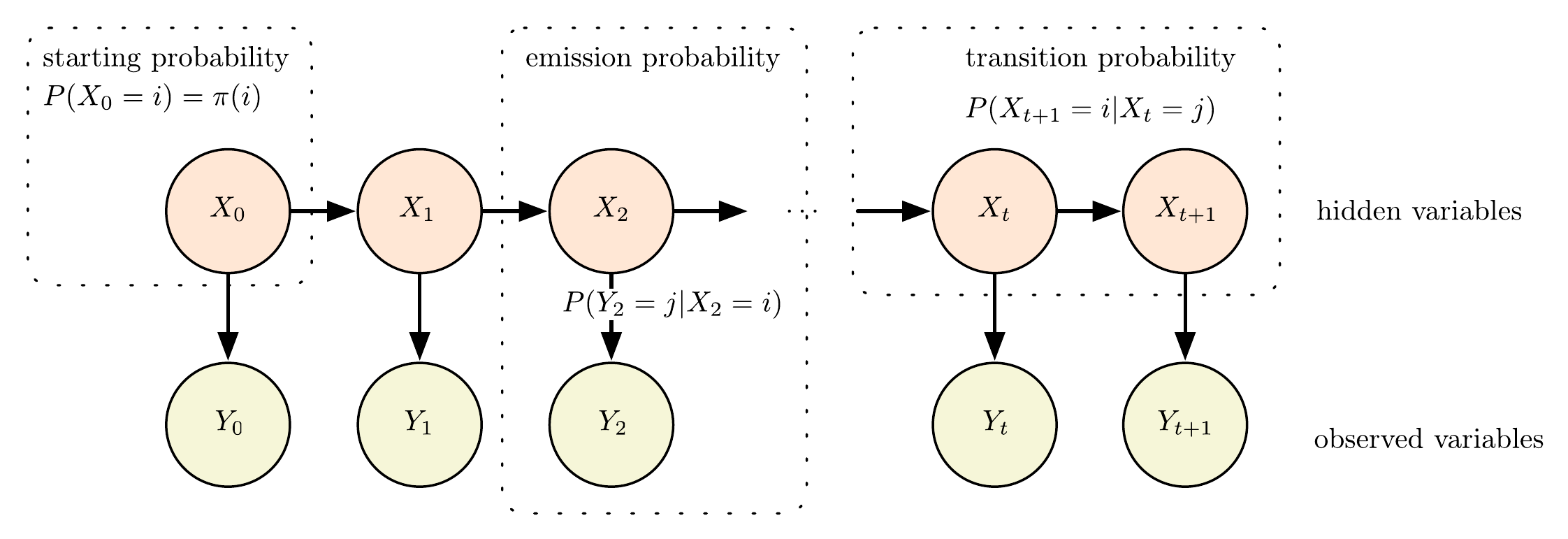}
\caption{Example of Hidden Markov Model}
\label{HMM_model}
\end{figure}

The HMM (Hidden Markov Model) consists of two types of random variables. The observed and the hidden variables. In figure \ref{HMM_model}
$X_i$ represents the hidden random variables and $Y_i$ the observed random variables. As you can see, $Y_i$ is independent to $Y_j$, $i\neq j$ and for $X_k$ we have the Markov property $P(X_{k+1}|X_k)$. $X_i \in S$ is a discrete random variable, where $S$ is the set of possible states of hidden random variable. We denote as $N=|S|$ the number of possible states of each hidden random variable.

The transition probabilities  $P(X_{k+1}|X_k)$ are usually stored in a rectangular matrix where $a_{ij}=P(X_{k+1}=S_i|X_k=S_j)$ and $S_i,S_j\in S$. The initial probability distribution $P(X_0)$ is called prior probability and is stored in a vector $\pi(i)=P(X_0=S_i)$. In figure \ref{HMM_gathered} we gathered this information in a table.

The observed variable can be either discrete random variable or continuous random variable. The probability that a hidden random variable
$X_k$ emitted an observation $Y_k=O_k$ is called emission probability. Please note that $O_k$ is the observation at time $k$. The observation might be a class, an integer, a real value and as we will see later, even a histogram. Depending the observation the observer random variable is either discrete or continuous. For example, imagine the stock market as a Markov process where the observed variable is the price of a share. The observed random variable is continuous and a common way to model such variables is using Gaussian distribution $N(\mu,\sigma)$. 
The emission probability $b_j(O_k)=P(Y_k=O_k|X_k=S_j)$ is a probabilistic function that models this probability.

\begin{figure}[h!]
\centering
\rowcolors{1}{white}{tableShade}
\begin{tabular}{lp{12cm}}
$X_k$ & hidden random variable \\
$Y_k$ & observed random variable \\
$S$ & set of possible values of hidden random variable \\
$N$ & number of possible values of hidden random variable \\
$a_{ij}$ & Transition probability $P(X_{k+1}=S_i|X_k=S_j)$\\
$b_j(O_k)$ & The probability that a state $X_k$ at state $S_j$ emitted an observation $O_k$\\
$A$ & Matrix that contain transition probabilities $A(i,j)=a_{ij}$\\
$B$ & Function that represent the emission probability $B(i,j)=b_j(O_k)$. In the case of discrete observed random variable, this
function can be represented with a matrix.\\
$\pi(i)$ & Initial probability $P(X_0=S_i)$\\
$\theta$ & Parameters of the Hidden Markov Model (HMM) $\theta=(A, B, \pi)$\\
\end{tabular}
\caption{Notation of HMM (Hidden Markov Model).}
\label{HMM_gathered}
\end{figure}

\centertitle{Problems to solve with HMM}
Hidden Markov Models are very useful in solving the following three problems \cite{Rabiner1989} :

\noindent
\textbf{Probability of an observation: }
For a given sequence of observations $\mathbf{O}=O_0,\dots,O_t$ and a model $\theta=(A, B, \pi)$[TODO: Explain them]
we want to compute the probability that $P(\mathbf{O}|\theta)$, that is the probability of
the observation given the specific model.

\noindent
\textbf{Model learning: }
Given data we want to optimise the model parameters $\theta=(A, B, \pi)$, such that $P(\mathbf{O}|\theta)$
is maximised. More formally :
\[
\theta^\star=\arg\max_{\theta}{P(\mathbf{O}|\theta)}
\]

\noindent
\textbf{Most probable hidden state sequence: }
For a given sequence of observations $\mathbf{O}=O_0,\dots,O_t$ and a model $\theta=(A, B, \pi)$
we want to find the most probable sequence of states of hidden random variables, that is
the sequence of states that best explain the observations.

In this thesis we will mainly focus on the last type of problem. The problem we will try to solve
is to find the most probable chord sequence (latent/not observed variable) that explain the melody played (observation).

\centertitle{Learning the model's parameters}


\centertitle{Most probable hidden sequence and dynamic programming}

The most probable hidden sequence of a Hidden Markov model is also know as the "optimal" state
sequence, associated with the given observation sequence, given that the "optimality" criterion
(or objective function) is the probability of that sequence ($P(\mathbf{Q}|\mathbf{O},\theta)$, where 
$\mathbf{Q}={q_1,q_2,\dots,q_t}$ is the hidden state sequence,$\mathbf{O}={O_1,O_2,\dots,O_t}$ is
the observation sequence and $\theta=(A,B,\pi)$ are the parameters of the model).

\begin{figure}[h!]
\centering
\includegraphics[width=0.6\linewidth]{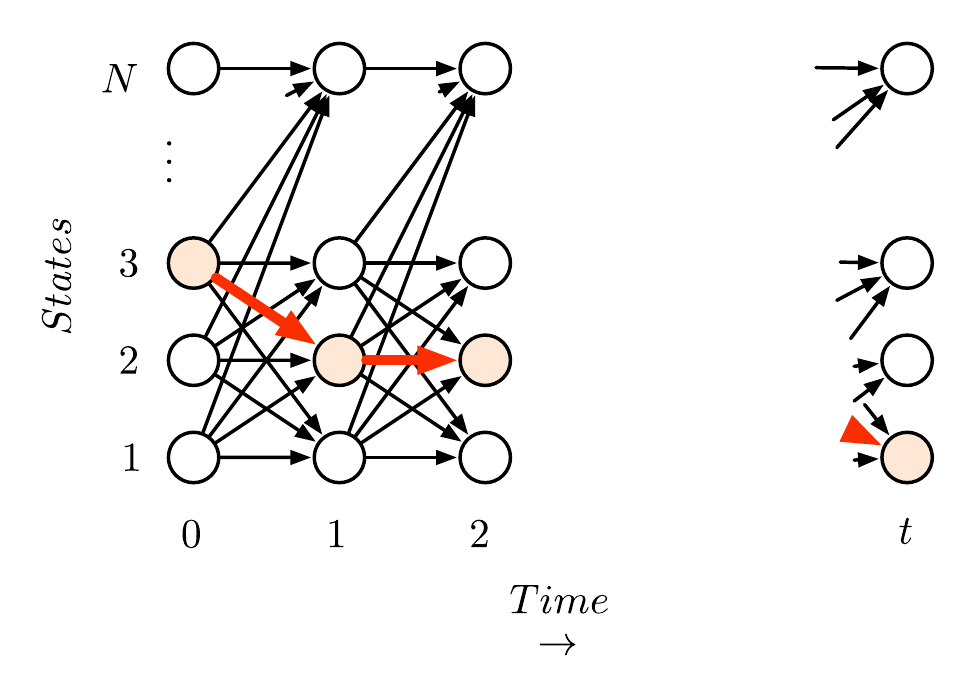}
\caption{Trellis view of Viterbi algorithm}
\label{HMM_viterbi}
\end{figure}

A trivial solution to that problem might be to compute the probability of each possible sequence.
It's easy to see that the number of possible paths increase exponentially to time. For example
of a state set of size $N$ and an $K$ observations the number of possible paths is $K^N$.
In figure \ref{HMM_viterbi} you can see a generic example of an HMM (Hidden Markov Model). The highlighted
nodes are the "optimal" states and the path in red is the "optimal" path (also known as Viterbi path).

A more feasible and trackable solution to this problem was given by Andrew Viterbi in late 60s \cite{Forney}. The algorithm
we will present you make use of Bellman's optimality principle \footnote{Bellman's principle of optimality --
The globally optimum solution includes no suboptimal local decision.} for the design of a dynamic programming algorithm.

\textbf{Algorithm}
\noindent

There are several variations of the algorithm, however in this thesis we aim to use a variation that is similar to an algorithm
know as forward algorithm.

The dynamic programming recurrent relations are:
\begin{align}
V_{0,k}&=P(O_0|X_0=S_k) \cdot \pi_k \label{viterbi_1} \\
V_{t,k}&=P(O_t|X_t=S_k) \cdot \max_{i}(a_{i,k} \cdot V_{t-1,i}) \label{viterbi_2}
\end{align}

$V_{t,k}$ represents the probability of the most probable state sequence until time $t$ for $S_k$ as final state. However,
what is of our interest is the path and not only the probability. For this we use the trick to save pointers to parent states in
the $\max$ step of second equation. The matrix $pa(k,t)$ contains the parent of state $S_k$ at time $t$.

The retrieval of the Viterbi path for an observation of length $T$ is computed using the following recurring formulas:

\begin{align}
y_T=&\arg\max_i(V_{T,i})\\
y_{t-1}=&pa(y_t, t)
\end{align}

\textbf{Complexity}
\noindent

In the computation of the probabilities we can see that initial step \ref{viterbi_1} takes time $O(N)$ since we have to initialise $V_{0,k}$
for all $|S|=N$ possible states. The second step \ref{viterbi_2} needs $O(N^2)$ time since we have to update
$V_{t,k}$ for all $N$ states and for each update we have to compute the maximum quantity $\max_{i}(a_{i,k} \cdot V_{t-1,i})$.
Also, by noting that this step will run for every previous time ($=T$) the overall complexity is $O(T\cdot N^2)$. As you can
see Viterbi algorithm improved the complexity from $O(K^N)$ to $O(T\cdot N^2)$.


To close this brief description of the background theory we will introduce you to the concept
of Variable Order Markov models which will be the component that makes our approach and system different
to the current state of art. As we will see, this model allows us to capture repetitive patterns
of chords and enhance the predictive power of our model.


\section{Variable Order Markov Model}

In this section, we introduce an extension of the fixed order Markov chain, the Variable Order Markov model (VOM) in order
to increase the prediction power of the system. The input will be a sequence of events and we aim to predict the next probable
event.


As we saw in the previous chapter, it is possible to model dependencies between chords and notes by using Bayesian networks. We will now see another probabilistic model which has been used in interactive music system of Francois Pachet, the Continuator.  \cite{Pachet2003}

In his paper, Pachet, described a system which has the ability to provide continuations of the musician's input. Continuation of melody is a prediction of the future of the melody which a musician played. It's easy to see that a simple 2 order Markov model can't be used for this task. We need a model which captures higher dependencies. Based on this idea, Pachet utilized Variable Order Markov models (VOM) so as to generate continuations. Let's discuss this idea more formally.



A learner is given a sequence $\textbf{q}=\{q_1,q_2,\dots,q_n\}$ where $q_i\in\Sigma$ and $q_iq_{i+1}$ is a concatenation of $q_i$ and $q_{i+1}$. The
problem is to learn a model $\hat{P}$ which will provide a probability of a future outcome given a history. 

More formally: 
\[
\hat{P}(\sigma | s) \text{-- the probability of } \sigma \text{ given a suffix } s
\]
where $s\in\Sigma^*$ is called "context" and $\sigma\in\Sigma$. $\Sigma$ is a finite alphabet (in our case this might be the possible notes or chords). The model $\hat{P}$ can be seen as the conditional distribution probability. Observe that the "context" can be, for example, a melody previously played and $\sigma$ the note which will be played next.  Let's assume that the "context" is a melody (note sequence). Then, the prediction of the next note given a sequence of notes requires the computation of the probability of a note -- symbol $\sigma$ given a sequence $s$. Let's leave this for a while and discuss a bit more about variable order Markov models.

In 1997, Ron \emph{et al.} \cite{Ron1997} showed that a variable order Markov model can be implemented 
with a subclass of probabilistic finite state automata, the \emph{Probabilistic Suffix Tree (PSA)}. Those trees provide
a fast way to query a tree with a sequence $s$ and retrieve the probability $\hat{P}(\sigma|s)$. PSTs have been used extensively in protein classification \cite{Bejerano2001} due to their speed of construction and lower needs of memory than HMMs. Please note that PSTs should not be confused with suffix trees, although they share common properties. The main difference is that the PST is a suffix tree of the reverse training tree \cite{Bejerano2001}.

The time and space complexities as reported in \cite{Ron1997} are presented in figure \ref{pstcomplex}
\begin{figure}[h!]
\centering
\begin{tabular}{|l|l|}
\hline
Complexity & Description \\
\hline
$O(Dn^2)$ & time for construction (learning) \\
$O(Dn)$ & space  \\
$O(D)$ & time for the query \\
\hline
\end{tabular}
\caption{time-space complexity of PST}
\label{pstcomplex}
\end{figure}
( D is the bound of the maximal depth of the tree ). Although it seems a bit restrictive the fact that the maximal order of the VOM is bounded, 
it is justified by the fact that it is impossible even for an experienced musician to retrieve from memory infinite long sequences. However, in 2000, Apostolico \emph{et al.} \cite{Apostolico2000}, improved PST's complexities to $O(n)$ for learning and $O(m)$ for predicting a sequence of length $m$.



\begin{figure}[h!]
\centering
\includegraphics[width=240pt]{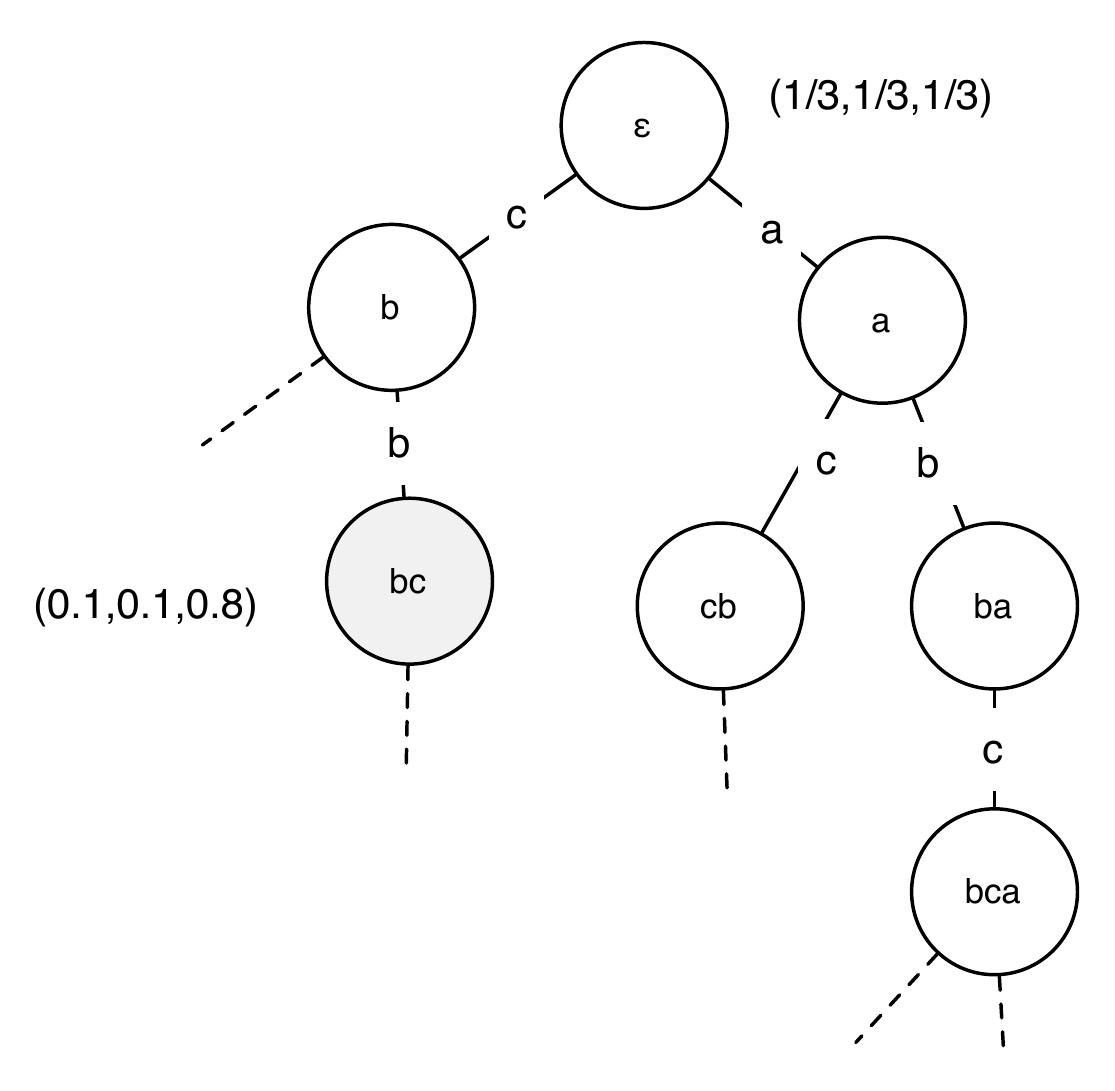}
\caption{
example of PST tree for $\Sigma=\{a,b,c\}$ -- Each node is labeled after the nodes we meet when we follow the path to the root. Let's assume we want to find the probability of $\hat{P}(c|"bc")$. 
Then, after querying the tree with the reverse of the string "bc" we end to the highlighted node. Thus, the probability is $0.8$. If we wanted to sample the next symbol of the sequence "bc", we should just use the vector $(0.1,0.1,0.8)$ as probability distribution for the symbols $a,b,c$.
}
\vspace{-17pt}
\end{figure}

\subsection{The Continuator}
Inspired by Ron's PSTs, Pachet simplified the data structure so as to be able to provide continuations. His approach is different than Ron's since the tree keeps information for all the played sequences (lose less). What is more, instead of computing the probability for each symbol, a pointer to the position of the initial training sequence is saved. That way, we are able to have access to several other aspects of the sequence's element, such as \emph{velocity} or \emph{duration}. 

\begin{algorithm}
\caption{learn sequence(s)}
\hfill\par
\begin{tabbing}
\textbf{F\textbf{\textbf{or}}} i\==1 \textbf{To} length(s)+1 \\
\> $T \leftarrow$ Tree root\\
\> \textbf{F\textbf{\textbf{or}}} j\==i \textbf{Down} \textbf{To} 1 \\
\>\> \textbf{If} $T[s_j]$ \= $\neq$ NULL \textbf{Then}\\
\>\>\> $T.pointer\_list.append(i+1)$\\
\>\> \textbf{Else} \\
\>\>\> $T.insert(s_j, i+1)$ \\
\>\> \textbf{End} \textbf{If} \\
\>\> $T \leftarrow T[s_j]$\\
\> \textbf{End} \textbf{F\textbf{\textbf{or}}} \\
\textbf{End} \textbf{F\textbf{\textbf{or}}}
\end{tabbing}
\end{algorithm}

The construction of the tree has time and space complexity $O(n^2)$ since in worst case we will
have to insert all the sub-sequences ($O(n^2)$) of the string in our tree. The tree initially is empty.
$T[symbol]$ returns the child of the node $T$ which is connected with label $symbol$. If there is no
such transition then it returns NULL. Each node has a list of pointers $pointer\_list$ to the position
of the next symbol. For example for a sequence $abb$ we will have a path $root\rightarrow b\rightarrow a$ which
pointer list contain a pointer to the second $b$ symbol.

\begin{figure}[H]
\centering
\includegraphics[width=220pt]{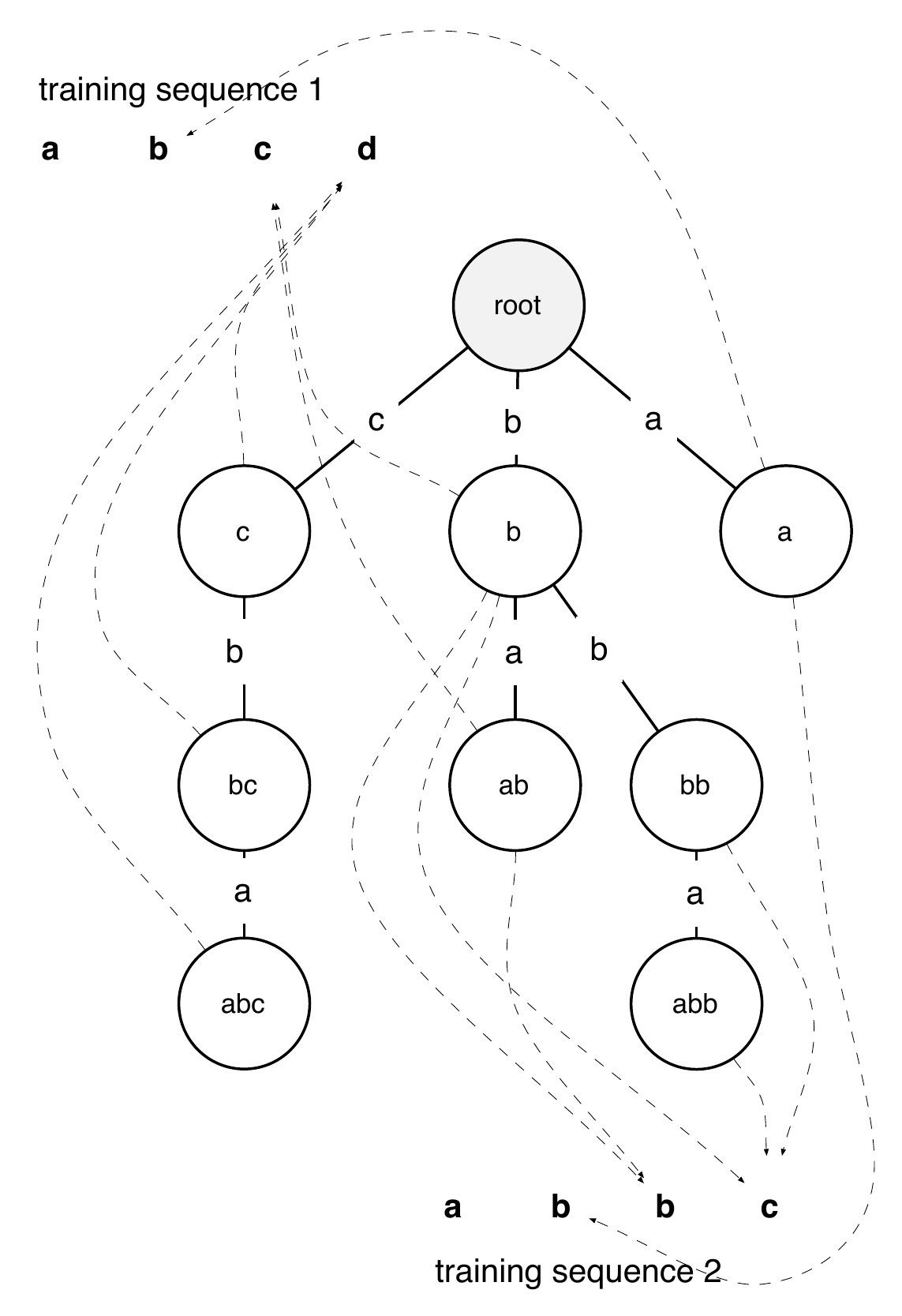}
\caption{
Result of learn\_sequence("abcd") and learn\_sequence("abbc").
}
\label{contree}
\end{figure}

In figure \ref{contree} you can see such a tree constructed from the sequence $abcd$ and $abbc$. The dashed arrows
show the pointers to the position in the learning sequence. So as to predict the next symbol
of a sequence we have to traverse the tree with the reverse of the query string and then sample uniformly at 
random a pointer from the pointer list. For example for the query $ab$ we search $ba$ in the tree and then sample a pointer from the pointer list (c from 1st sequence or b from 2nd sequence) with equal probability.

\chapter{Design and implementation}

\epigraph{\emph{Music gives a soul to the universe, wings to the mind, flight to the imagination and life to everything.}}{Plato}

\section{Anticipation and surprise in music improvisation}

A crucial aspect of a jam-session musician is the anticipation. That means that jam-session support musician should
be able to predict musical events and provide a musical stability in the performance. For example, a musician
responsible for chord accompaniments must be able to understand patterns in music of the improvisation
and try to predict and protect the stability of this structure. That way the soloing/improvising musician
will depend on the assumption that the structure is stable and thus will be able to improvise in harmony.

Our observation was that during a jam rehearsal there is a convergence of a stability as the time pass. Initially
there is no preparation and thus no structure established. However, all the musicians tend to introduce ideas 
and variations that might enter the theme or forget them. Depending  his/her role in the rehearsal, the musician
will either be the one that introduce the innovation or the one that will try to provide accompaniments. The goal
of this thesis is to develop a system that will serve the duties of the second. 


The input of the system if a musician that improvise polyphonic or monophonic melodies. Without any prior
knowledge about the music played the task is to predict next chords and if possible converge to an underlying structure. This task has several
not-so-obvious problems that we must solve. One of the greatest problems is that we don't have any information about the chords 
that have been played. Thus, our system must have a subsystem that will be able to infer chord sequences
according the history of played music. 

Another problem is that we have to predict next chords using information learned during the rehearsal
(online learning). For example, in 12-bar blues the structure (chord progression) is strict and repeated every 12 bars. However, an improvisor
who plays on 12-bar song will not play strictly the same melody but will try to play in harmony with the chords. How can we 1)find the chord sequence that fits the history and 2)understand the structure and predict next chord? Those two questions are answered using two model that we have already
discussed in previous chapter.

\section{Our method}
 
The system designed and tested in this thesis consists of two subsystems. The inferencer and the predictor. Before we describe each
subsystem separately we will show you the interaction of those two subsystems.

\begin{figure}[H]
\centering
\includegraphics[width=0.8\linewidth]{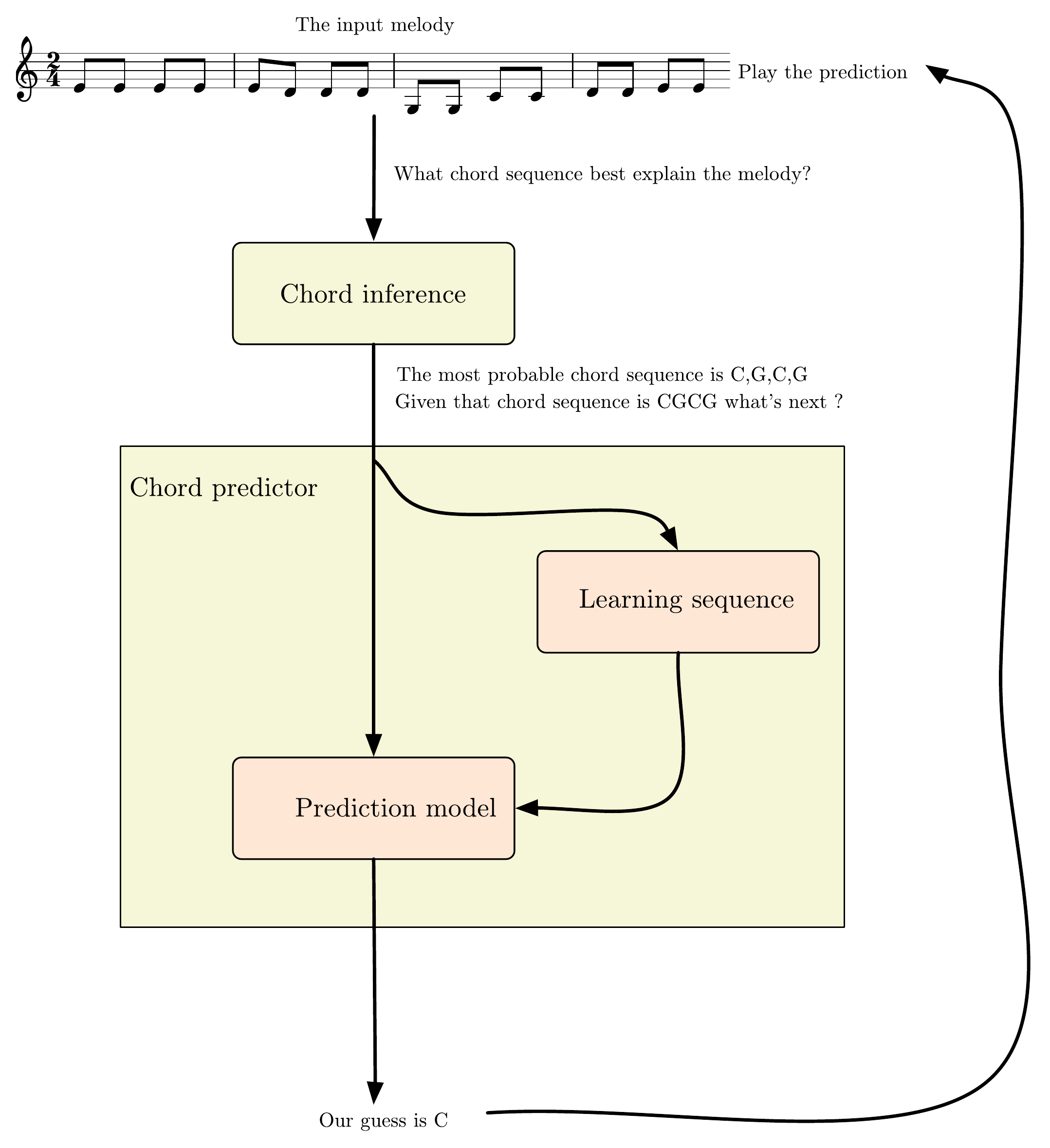}
\caption{Brief system overview}
\label{brief_overview}
\end{figure}

As we can see in figure \ref{brief_overview} the system initially takes as input the melody played by a soloing musician. As you can 
see there is no knowledge about the chord or the structure. The inferencer using knowledge learned from corpus try to find 
a chord sequence that explain the melody. That means that this chord sequence might be in the musician's mind or played from other
peer musicians and we have no way to interact with them. The inferencer outputs the sequence and feeds that sequence to the predictor.
The predictor then use that sequence to learn the patterns and add this knowledge to the chord predictor. The chord predictor
then using the chord sequence will try to predict next chord. As you can see, our aim is to improve the prediction using knowledge
learned from the inferred sequence.

One key limitation to our design choices is that each subsystem must work fast enough so as to be able to work in real-time. That
means that the models and algorithm that we choose should be run very fast in a descent computer. Also, so as to simplify the task we
made the following assumptions:
\begin{enumerate}
\item Tempo is fixed. That means that it doesn't depend on the performance.
\item Chord changes occur at bar change. That is that each bar is assigned one and only chord.
\item There doesn't occur any key-modulations. That means that the key remains constant during the performance. The recognition
of key changes is a research topic on itself.
\end{enumerate}

\subsection{Chord modelling}
Chords modelling and representation is of great interest on this thesis since this decision affected mostly the performance and the output of the system. We will describe 3 representations that we tried and what was the drawback of each representation.

The first idea was to model each chord as a 12-vector followed by the pitch of the root. Each position of this vector represents the note relative to the root that is played. For example for a Cmajor chord we have that notes at position relative to the root $0, 5, 7$ are played, thus Cmajor will be represented as $(C,<1,0,0,0,0,1,0,1,0,0,0,0>)$. It's easy to see that this representatios is very "hard". That means that for similar chords (e.g. Cmajor and Cmaj9) we will have different representations. This "hardness" of this representation incresed the "zero-frequency" problem since most of the songs had several variations of chords. That happens because usually musicians have a basic chord structure on their mind and then they introduce several stylistic note into the chords. The second representation we will see tries to deal with this problem.

In out approach we decided to include musical knowledge into chord representation and simplify chords in a way that similar chords has the same representation. But what is the similarity threshold? We decided to split the chords according the quality of the 3rd and 5th interval. That way we separated chords to 5 chord types:

\begin{figure}[h!]
\centering
\begin{tabular}{l|l|l}
Chord type & Third interval quality & Fifth interval quality \\
\hline
Major & major & perfect \\
Minor &  minor & perfect \\
Augmented & major & augmented \\
Diminished & minor & diminished \\
Suspended & none (2nd or 4th instead) & perfect
\end{tabular}
\caption{chord types}
\end{figure}

Thus we have for each of 12 roots, 5 different chord types, thus 60 different chords in total. As we will see, that chord simplification is enough so as to train and make our system functional.

This issue has been also studied and researched in literature \cite{Paiement2009}.

\subsection{Chord inference: Hidden Markov model}

As we already saw this subsystem is responsible in answering the question "What is the chord sequence that explains the melody played?".
In Background chapter we saw a specific model and algorithm that answers that question. Hidden Markov models is the model we chose for this task due to their simplicity, time complexity and because they are most appropriate model for answering that kind of question. So as to apply Hidden Markov models (HMM) on this task we have to set the question in a probabilistic context. 

We have two kinds of random variable sequences. The note sequence that is fully observable from the system and the chords that are latent (not observable). We denote $\mathbf{O}={O_0,O_1,\dots,O_T}$ the sequence of the random variables that corresponds to observable melodic level ($O_0$ is for bar $0$, etc) and with $\mathbf{C}={C_0,C_1,\dots,C_T}$ to the hidden random variable sequence that correspond to the chord level. 
\[
P(\text{structure}|\text{surface})=P(\text{chord sequence}|\text{melody})
\]

An interesting question that we had to answer was what will be an appropriate representation for the melodic level's random variables ($\mathbf{O}$). The answer was inspired by MySong system \cite{Simon2008} which is a system developed by Microsoft for harmonisation (chord inference) of vocal tracks. The key observation is that each chord supports melodies that have specific distribution of it's notes in each pitch class. That means that a Cmajor chord will support melodies that have most of their notes in C pitch class some of them also in G and almost none in D\# or G\#. Thus, we can assign each chord with a distribution of pitch classes. Those distributions can be easily learned from appropriate data set.

\begin{figure}[h!]
\centering
\includegraphics[width=0.5\linewidth]{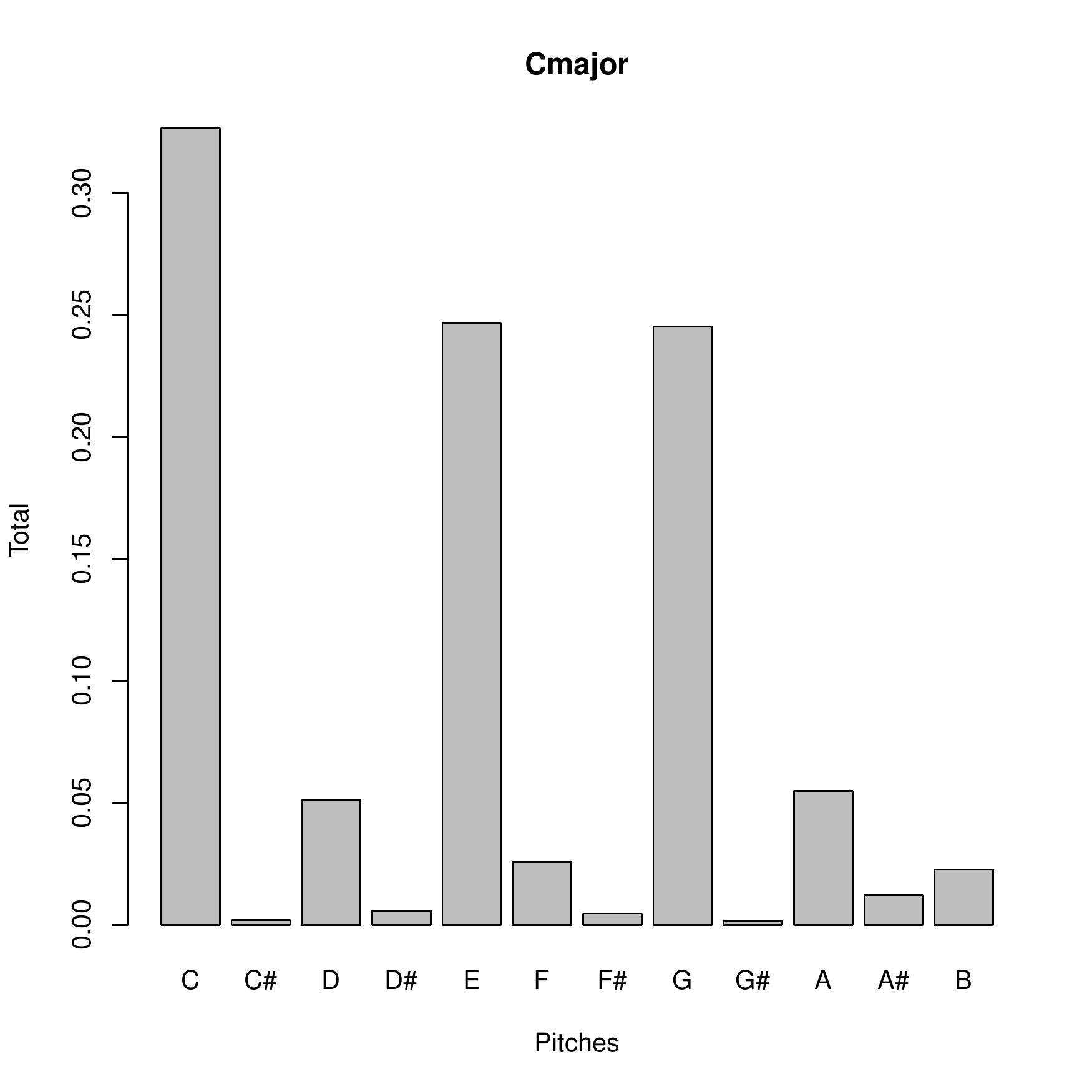}
\caption{Cmajor pitch discrete distribution}
\label{Cmajordist}
\end{figure}

\subsubsection{Emission probabilities}

As we saw , we map each chord with a discrete distribution of pitch classes. This represents
how much of each note stayed in a pitch class during the playback of this chord. Although we will give details
about training and data set later we will give a description of the way we learn those distributions.


We denote with $\mu_c$ the 12-demensional vector that correspond to the pitch class distribution of chord with index $c$.
That means for the chord with index $1$, $\mu_1$  represents the percentage of time that melody spent in each pitch class.

e.g \[\mu_1=(0.43,0.2,\dots,0.001)\]

\noindent
Also, $\mu_{i,j}$ is the percentage of time melody spent in pitch class $j$ for chord with index $i$.

\noindent
The following process is followed so as to learn $\mu$. For each bar we initialise a temporary 12-dimensional vector $t=(0,0,0,\dots,0)$. Each element $t_i$ of this vector represents
the percentage of time that a note spent in the corresponding $i$-th pitch class. The following process compute that 
vector. For each note in the bar we extract the octave information so as to assign it to a specific pitch class (e.g. C3 to C pitch class).
Then we add the note duration to it's corresponding vector position. For example for a half note of C3 we increase $t_0$ by $4$.
Next, we normalise the profile so as avoid problems that might occur by the change of the time signature (although we assumed that
this won't happen that way we are safe in case we remove that assumption).
Then we update $\mu_{c,j}=\mu_{c,j}+pc_j$ where $c$ is the index of the chord that accompanied that bar.
When the training finish we normalise again $\mu_{c,j}$ so as $\sum_j{\mu_{c,j}}=1$. That way $\mu_i$ corresponds to a probability distribution. 

So far we've seen a way to learn the distributions from data however we need a way to model emission probabilities - that is
the probability a chord have accompanied the specific observation (group of notes).

\begin{figure}[h!]
\centering
\includegraphics[width=0.7\linewidth]{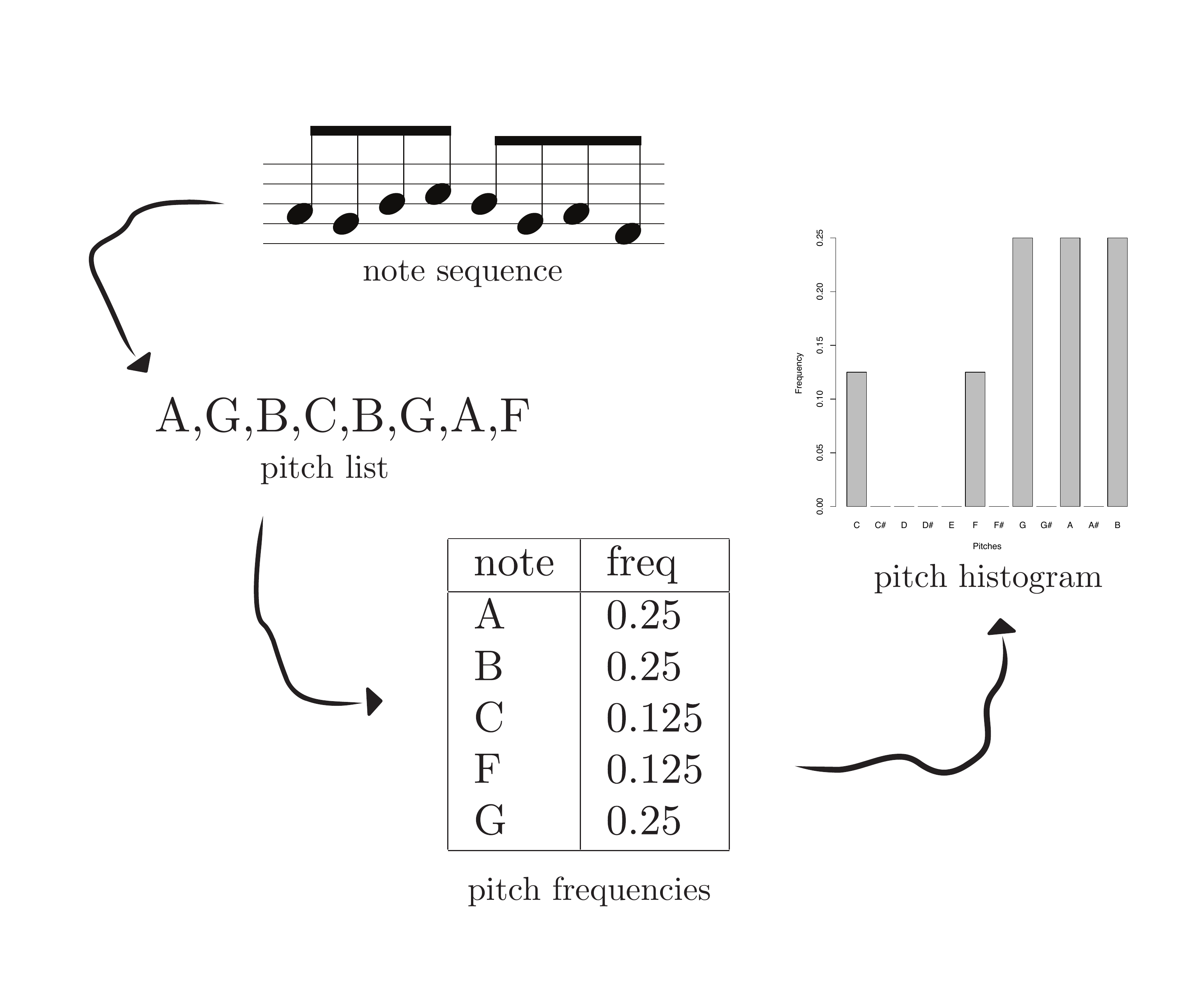}
\caption{Note to histogram process}
\label{note2histrogram}
\end{figure}

We denote with $h$ a 12-dimensional vector which contains the pitch-class distribution of the melody of the observed bar.
The computation of $h$ occur with the same process we used to learn $\mu$ (figure \ref{note2histrogram}). Thus now we have to find a way
to compute $P(melody|chord)$. It's easy to observe that the probability of a note $n$ played given a chord $c$ is \[P(n|c)=\mu_{i,pc(n)}\] where $pc(\cdot)$ is a function that maps a note to it's pitch-class. For a bar $bar$ a set of notes, the probability that it was accompanied by a chord with index $i$ is 
\[
P(bar|c)=\prod_{j=1}^{T}{P(bar_j|c)}=\prod_{j=1}^{T}{ \mu_{c,pc(bar_j)}}
\]
where $bar_j$ is the $j$-th note of $bar$ and $T$ it's total number of notes. So as to avoid small probabilities and underflows we will work with log probabilities instead. Thus
\begin{align}
\log{P(bar|c)}=\sum_{j=1}^{T}{\log{P(bar_j|c)}}=\sum_{j=1}^{T}{ \log{\mu_{c,pc(bar_j)}}} \label{eq:tmp1}
\end{align}

However we notice that notes that are in the same pitch class will produce the same probability $\log{\mu_{i,pc(bar_j)}}$. If we count the number
of notes in each pitch class and we assign them in a vector $t_i$ that contains such information we can replace the equation \ref{eq:tmp1} with
\begin{align}
\log{P(bar|c)}=\sum_{j=1}^{12}{t_j \cdot \log{\mu_{c,j}}} \label{eq:tmp2}
\end{align}

Also note that $h_i = t_i/T$ thus $t_i=h_i \cdot T$. By replacing this to \ref{eq:tmp2} we have

\begin{align}
\log{P(bar|c)}=\sum_{j=1}^{12}{t_j \cdot \log{\mu_{c,j}}} = \sum_{j=1}^{12}{h_j \cdot T \cdot \log{\mu_{c,j}}} \propto \sum_{j=1}^{12}{h_j \cdot \log{\mu_{c,j}}} \label{eq:tmp3}
\end{align}

To summarise, for a bar $bar$ we extract it's pitch-class distribution and we assign it to a 12-dimensional vector $h_i$. Then, to compute the probability the a bar is accompanied by a chord with index $i$ we compute the dot-product of $h$ with the logarithm of $\mu_i$.
\begin{align}
\log{P(bar|c)} \propto h \cdot \log{\mu_c} \label{eq:emission}
\end{align}

Now we have to find a way to represent chords and learn the transitions between them.

\subsubsection{Transition probabilities}

As we described, Hidden Markov models need emission probabilities and transition probabilities to be specified so as to work. In our 
case we have already seen emission probabilities and now we will see how to compute $P(chord_{t+1}|chord_{t})$.

Initially we have to find an appropriate representation of the chords. That was one of the biggest and most difficult
decision we had to make in this thesis since the whole system could suffer by a bad representation. As we saw in the Background
chapter the chord is a set of notes that played simultaneously. Recalling from music theory, chords in western music consist of the root note,
the 3rd interval note and 5th interval note. For example a major chord consists of the root note, a minor 3rd (that is 4 semitones) and a perfect fifth (7 semitones). According the quality of the 3rd and 5th interval of the chord we distinguish 5 types of chords (minor, major, diminished, suspended, augmented).

So as to represent a chord we followed this simple representation. Each chord is named by it's root name followed by the type of chord  e.g. Cmajor. As we will see in the experimental results, the number of different types of chords that we allow affected significantly the performance of our system. So as to proceed with our explanation we assume that we have only two types of chords - minors and majors. Thus 
\[
chord_t \in Root \times \{minor, major\}
\]
where $Root=\{C,C\#,D,D\#,E,F,F\#,G,G\#,A,A\#,B\}$. Each $chord_t$ is an integer that correspond to a chord at bar $t$. The mapping will be described later.

As we know, the hidden level of a Hidden Markov model (HMM) is a Markov model of order 1. That means that $P(chord_{t+1}|chord_t)$ 
can be specified by a $m\times m$ transition matrix $A_{ij}$ , where $m$ is the number of chords, which contains the probability that a chord with index $i$ will transit to a chord with index $j$. Give chord sequences extracted from data set those probabilities
can be learned by simple counting.

\begin{align}
P(chord_{t+1}=i|chord_{t}=j)=A_{ij}=
\frac{
\# \text{ of transitions from } i \text{ to } j
}{
\# \text{ all transitions} = |\text{training sequence}|-1
} \label{eq:transition}
\end{align}

\begin{figure}[h!]
\centering
\includegraphics[width=1\linewidth]{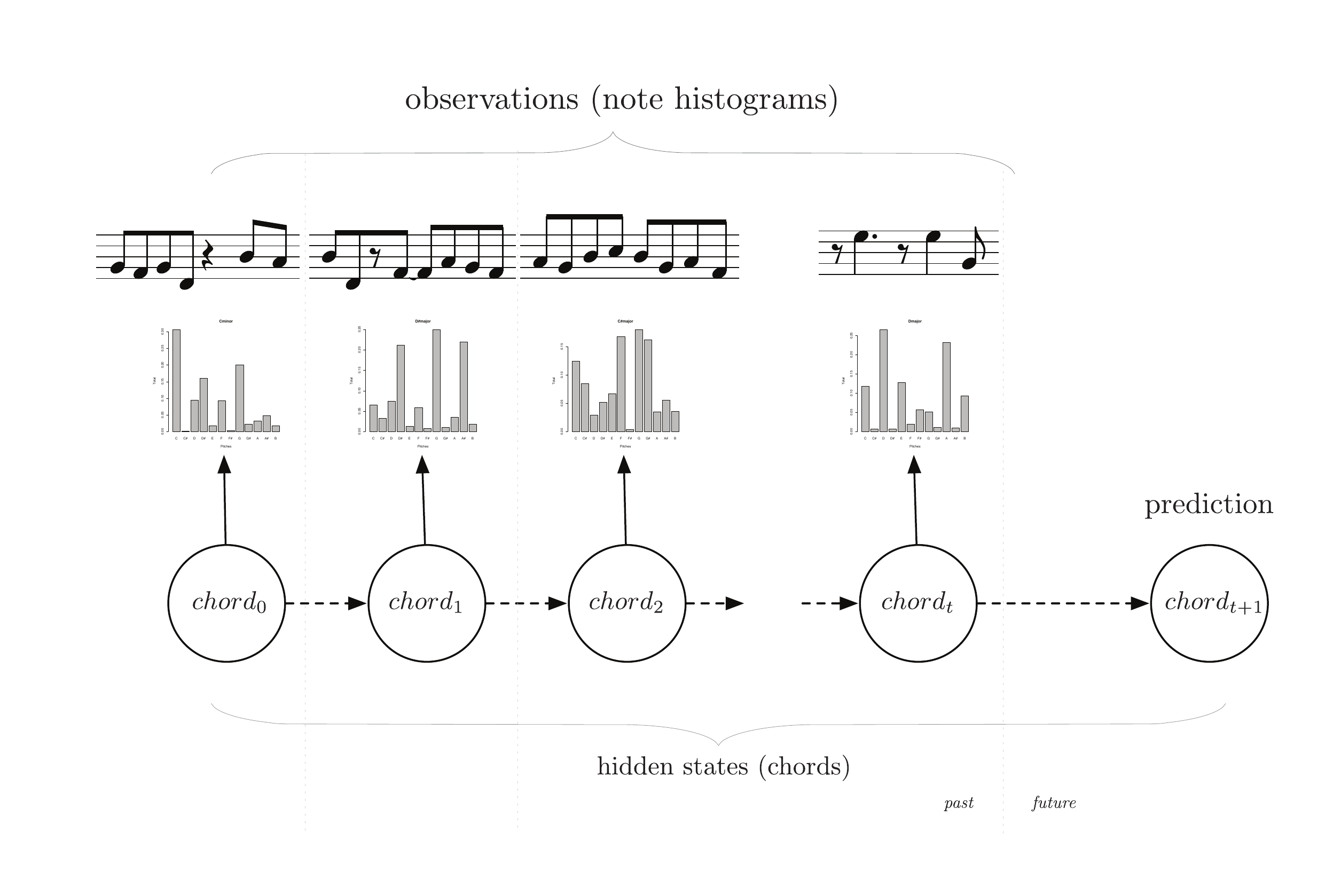}
\caption{HMM for chord inference}
\label{HMM_for_chord_inference}
\end{figure}

\subsubsection{Initial probabilities}

The only part missing is the initial probabilities. That is the probability of chord $i$ to be the first chord in the sequence $P(chord_0=i)$. This
probability can be easily learned from training set
\[
\pi_i=P(chord_0=i)=\frac{\#\text{ sequences starting with chord i}}{\# \text{ of training sequences }}
\]

\subsubsection{Chord inference}

To summarise our inferencer the topology of Hidden Markov model can be seen in figure \ref{HMM_for_chord_inference}.
At time $t$ we assign the pitch-class distribution of the bar to observation $o_t$ and the emission probability is
\[
\log{P(o_t|chord_t)}= \alpha + h \cdot \log{\mu_c}
\], where $\alpha$ is a normalising factor, and the transition probabilities 
\[
P(chord_{t+1}=i|chord_{t}=j)=A_{ij}=
\frac{
\# \text{ of transitions from } i \text{ to } j
}{
\# \text{ all transitions} = |\text{training sequence}|-1
} \label{eq:transition}
\]

Our task is now to infer the most probable chord sequence $\mathbf{C}=\{chord_0,chord_1,\dots,chord_t\}$ for a history of observations $\mathbf{O}=\{o_0,o_1,\dots,o_t\}$. That is 
\[
P(\mathbf{C}|\mathbf{O})
\]

We have already seen an algorithm that allows us to answer such questions very fast -- \textbf{Viterbi} algorithm.
The objective function that we like to maximise is 
\[
L=\log{P(\mathbf{C}|\mathbf{O})}
\]

However, recalling Bayes theorem we have that $P(\mathbf{C}|\mathbf{O})=\frac{ P(\mathbf{O}|\mathbf{C}) \cdot P(\mathbf{C} )}{ P(\mathbf{O}) } \propto  P(\mathbf{O}|\mathbf{C}) \cdot P(\mathbf{C} )$. Thus, $\log{P(\mathbf{C}|\mathbf{O})} \propto \log{P(\mathbf{O}|\mathbf{C})} + \log{P(\mathbf{C} )}$ and by maximising the right part we maximise the left part too. The objective function becomes:
\[
L= \log{P(\mathbf{O}|\mathbf{C})} + \log{P(\mathbf{C} )}
\]
and the task:
\[
\mathbf{C^*}=\arg\max_{\mathbf{C}}{(\log{P(\mathbf{O}|\mathbf{C})} + \log{P(\mathbf{C} )}
)} 
\]

The recurrent equation of the Viterbi algorithm for this task are:

\begin{align}
V_{0,k} =& \log{P(O_o|C_o=k)} + \log{\pi_k} \\
V_{t,k} =& \log{P(O_t|C_t=k)} + \max_{i}{ \log{A_{i,k}} + V_{t-1,k}}
\end{align}

\noindent
after replacing $\log{P(O_o|C_o=k)}$ with emission probability formula (equation \ref{eq:emission}) we have:

\begin{align}
V_{0,k} =&  h_0 \cdot \log{\mu_k} + \log{\pi_k} \\
V_{t,k} =& h_t \cdot \log{\mu_k} + \max_{i}{ \log{A_{i,k}} + V_{t-1,k}}
\end{align}

\noindent
Finally as we described in Background chapter, we retrieve the Viterbi path using the equations

\begin{align}
y_T=&\arg\max_i(V_{T,i})\\
y_{t-1}=&pa(y_t, t)
\end{align}

\subsubsection{Biasing emission over transition}

Another idea that we adopted from MySong was to introduce a bias factor in the objective function of Viterbi. That way we can bias chords that have better 
emission probabilities that transition. The reason why want to do so is because that way we can control the variability of inferenced chords and thus of the predicted ones.
For example if we assign more importance on emission probabilities then the system will become more sensitive to the melody just played instead of the chord sequence that is most probable.

More formally, the way we paraneterise the importance is by introducing $\alpha$ factor in the objective function.
\[
L= (1-\alpha) \cdot \log{P(\mathbf{O}|\mathbf{C})} + \alpha \cdot \log{P(\mathbf{C} )}
\]
, where $\mathbf{O}={O_0,O_1,\dots,O_T}$ the sequence of the random variables that corresponds to observable melodic level ($O_0$ is for bar $0$, etc) and $\mathbf{C}={C_0,C_1,\dots,C_T}$ chord progression.

So far, we saw a method to infer the chord sequence that accompanies a given a history of 
pitch-class distributions. Next we will see the subsystem responsible for chord prediction.





\subsection{Chord prediction: Variable Order Markov model}

One of the limitations of 1-st order Markov models is that it can't catch higher order dependencies.  So as to illustrate
this limitation imagine a chord sequence that is Cmjaor, Gmajor, Cmajor, Aminor, Cmajor, Gmajor, Cmajor or using a
sorter notation C,G,C,Am,C,G,C. 
A 1-st order Markov model will give positive probability for the next chord to be either Am or G. However,
a closer look will reveal that there is a pattern (that is C, G, C, Am). How can we extend the 1-st order
Markov model so as to a) catch those dependencies and b) be as fast as possible to train and query ?

Variable Order Markov models (VoM) as we saw are models that are used in lose less compression and can be used also for
prediction, due to their connection to the first problem \cite{Begleiter2004}. Those models allows us to answer questions such us
what is the next symbol given a context ? Or more formally:
\[
P(s|\sigma), \text{ where } s \text{ is next symbol and } \sigma \text{ the context. }
\]

\noindent
In our task, the question is "Given a chord sequence, what is next chord ?". Or more formally:
\[
P(chord_{t+1}|chord_0^t), \text{ where } chord_0^t \text{ is the chord sequence until } t \text{ and } chord_{t+1} \text{ next chord }
\]

For this thesis, we implemented a Variable Order Markov model (VoM) using a tree structure which capture variable order
dependencies and allows us to
\begin{itemize}
\item Train the model on-line (during performance)
\item Train and query really fast (at real-time)
\end{itemize}

\begin{figure}[H]
\centering
\includegraphics[width=0.74\linewidth]{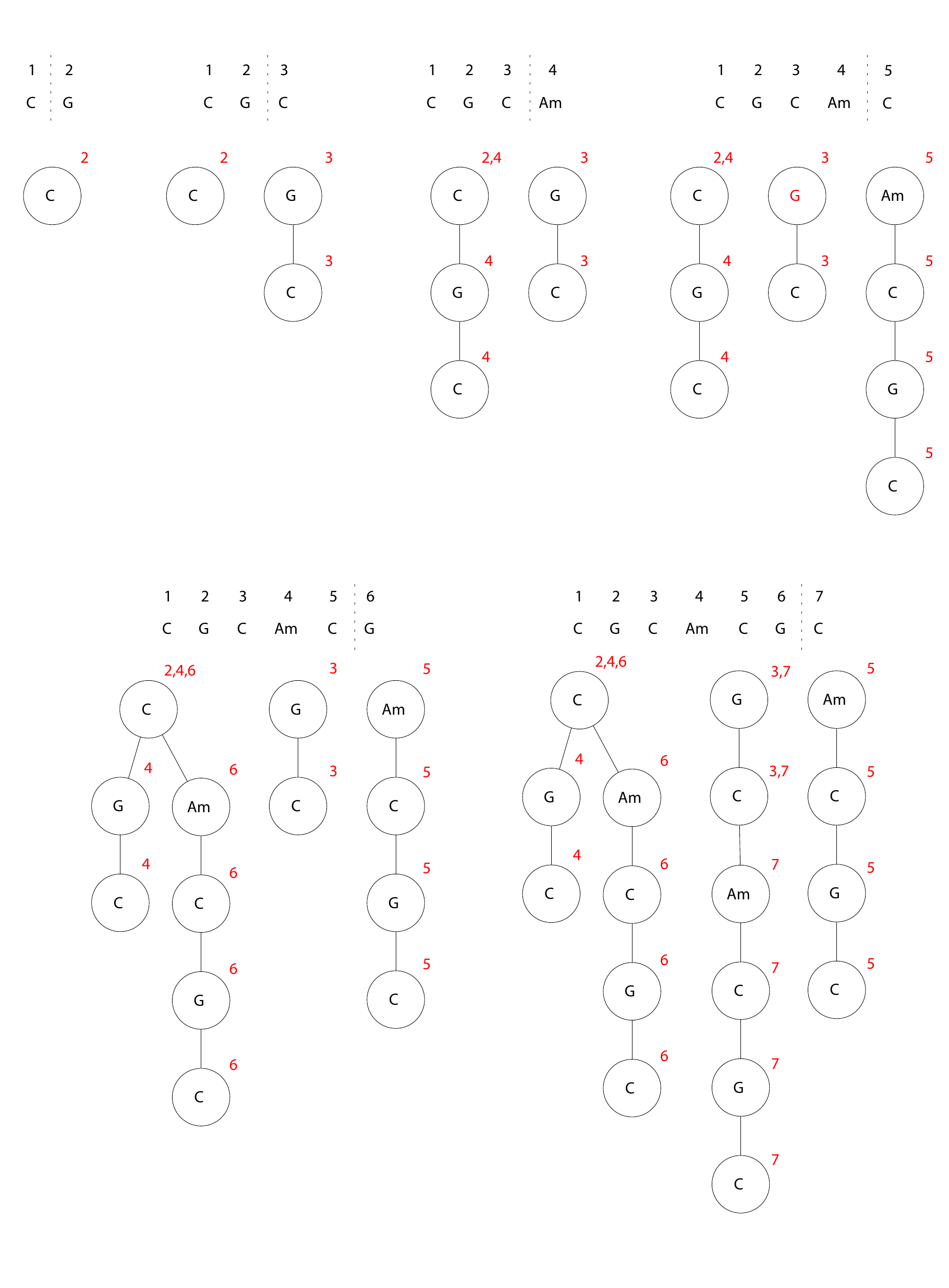}
\caption{Training of an empty Variable Order Markov model with chord sequence CGCAmCGC}
\label{vom_training}
\end{figure}

\subsubsection{Example}

Recalling the example we saw in the beginning of this section let's assume that we have the following chord sequence
\begin{center}
C,G,C,Am,C,G,C
\end{center}
and initially an empty tree. We will show how to train the tree and try to predict next symbol. As we can see in figure \ref{vom_training}
we construct the tree incremental. That is we start parsing the sub sequences from right to left. For example,
initially we start with symbol C preceding symbol G. Thus we create a node C with pointer to position 2 of the sequence. The numbers
in red are the pointers to the training sequence.

Given that we have trained our model we need to predict next chord. More formally, the question we want to answer is
\[
P(chord_{t+1}|chord_0^t=\text{CGCAmCGC})
\]

To answer this question we start from the rightmost symbol of our conditional sequence and travel the tree until the node we are
is leaf (that is, it doesn't have children). 
In our example we start with symbol C and we visit node G and the C (reverse order).
As we can see, the pointer of that node is 4 which means that next chord is Am with probability $P(chord_{t+1}=\text{Am}|chord_0^t=\text{CGCAmCGC})=1$.

\begin{figure}[h]
\centering
\includegraphics[width=0.5\linewidth]{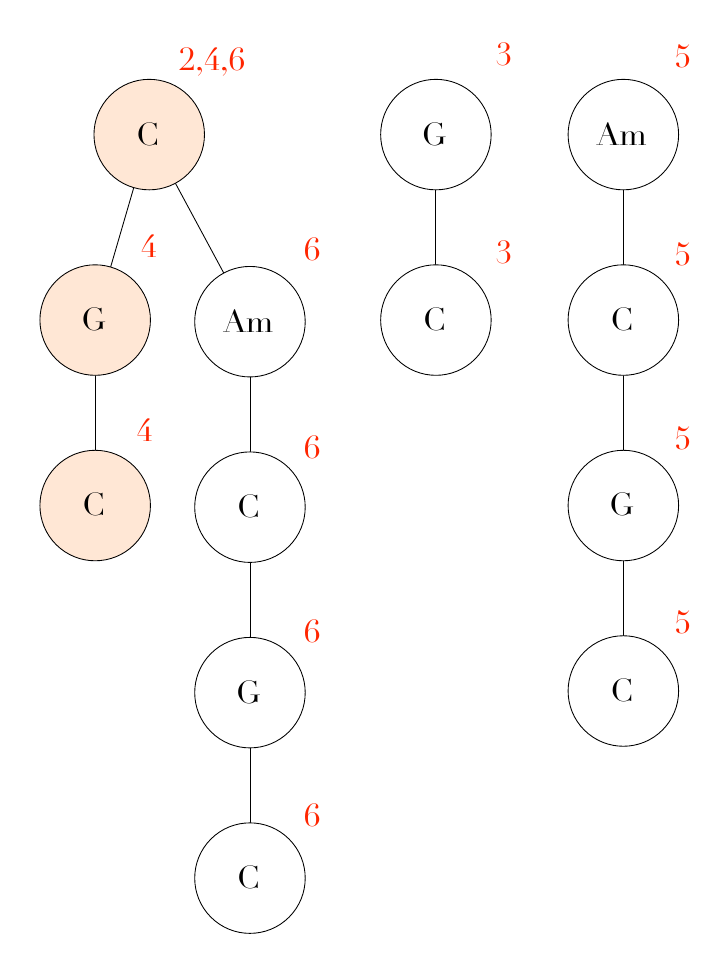}
\caption{Example of query}
\label{vom_query_example}
\end{figure}

What if we wanted to answer the question $P(chord_{t+1}|C)$ ? Again, from the tree we can see that there are three pointers 2,4,6. That means
that there is $\frac{1}{3}$ probability of each pointer to be the next chord. However you can see that pointer 2 and 6 point to chord G
and pointer 4 to Am. Thus, this gets translated to $\frac{2}{3}$ for G to be next chord and $\frac{1}{3}$ for Am. If we want to return the
chord that maximise the probability we will have to count each pointer endpoints and return the most pointed chord. The reason why
we use pointers instead of chords is because we might want to store more informations such as duration of the chord or stylistic options
such as adding a 9th interval. Thus, this decision improves the expandability of our system.

\subsection{Hybridation}

As we saw our system consist of an inferencer and a predictor. We have already describe how each component
work and we will proceed to the description of the main prediction algorithm.

Every time a bar finish, the system computes the 
discrete pitch class distribution of the melody and call prediction algorithm with that as input.
The prediction algorithm append the observation $x$ to an observation list which contains
the observation history until then. Using that history, it calls Viterbi algorithm and infer a chord sequence
that best explains that observation history. The chord sequence then is used to query the Variable order Markov
model (VoM) so as to fetch next possible chord (prediction). We are ready now to play that chord and
update the Variable order Model with the observation history.

\begin{algorithm}
\caption{Next\_chord\_prediction}
\textbf{input}: bar note histogram $x$ \\
\textbf{output}: next chord
\hfill\par
\begin{tabbing}
$append(observations, x)$ \\
$chord\_sequence$ $\leftarrow$ $viterbi(observations)$\\
$predicted\_chord$ $\leftarrow$ $VoM\_query(chord\_sequence)$\\
$output(predicted\_chord)$\\
$VoM\_train(chord\_sequence)$\\
\end{tabbing}
\end{algorithm}

\subsubsection{Zero frequency}
One of the problems we had to deal with was the case where the Variable order Markov Model (VoM) doesn't have information about the 
next chord. This is possible mostly at the beginning of each session were there are not enough data to train the system, or
the jam session is not very stable - that is, there is no structure or chord progression pattern. In that case, VoM would return the null
value, which means that no prediction can be made. This is problem is known in literature as `Zero frequency problem`.

Our strategy to resolve this situation was the following. In the event of a zero-frequency we will use the information learned
offline from the corpus. As we saw, this information is stored in the Hidden Markov Model in transition probabilities. This is
equivalent to having trained the Variable Order Markov model with the corpus using as maximum allowed order equal one. To summarise, query our tree for a prediction, and if this fails we search for the chord that maximise the probability 
\[
chord_{t+1}=\arg\max_i{ P(chord_{t+1}=i|chord_{t}=j) }
\]
where $j$ is the last played chord as inferenced by the Hidden Markov Model.   

\begin{figure}[h!]
\begin{center}
\hspace*{-40pt}
\includegraphics[width=1.2\linewidth]{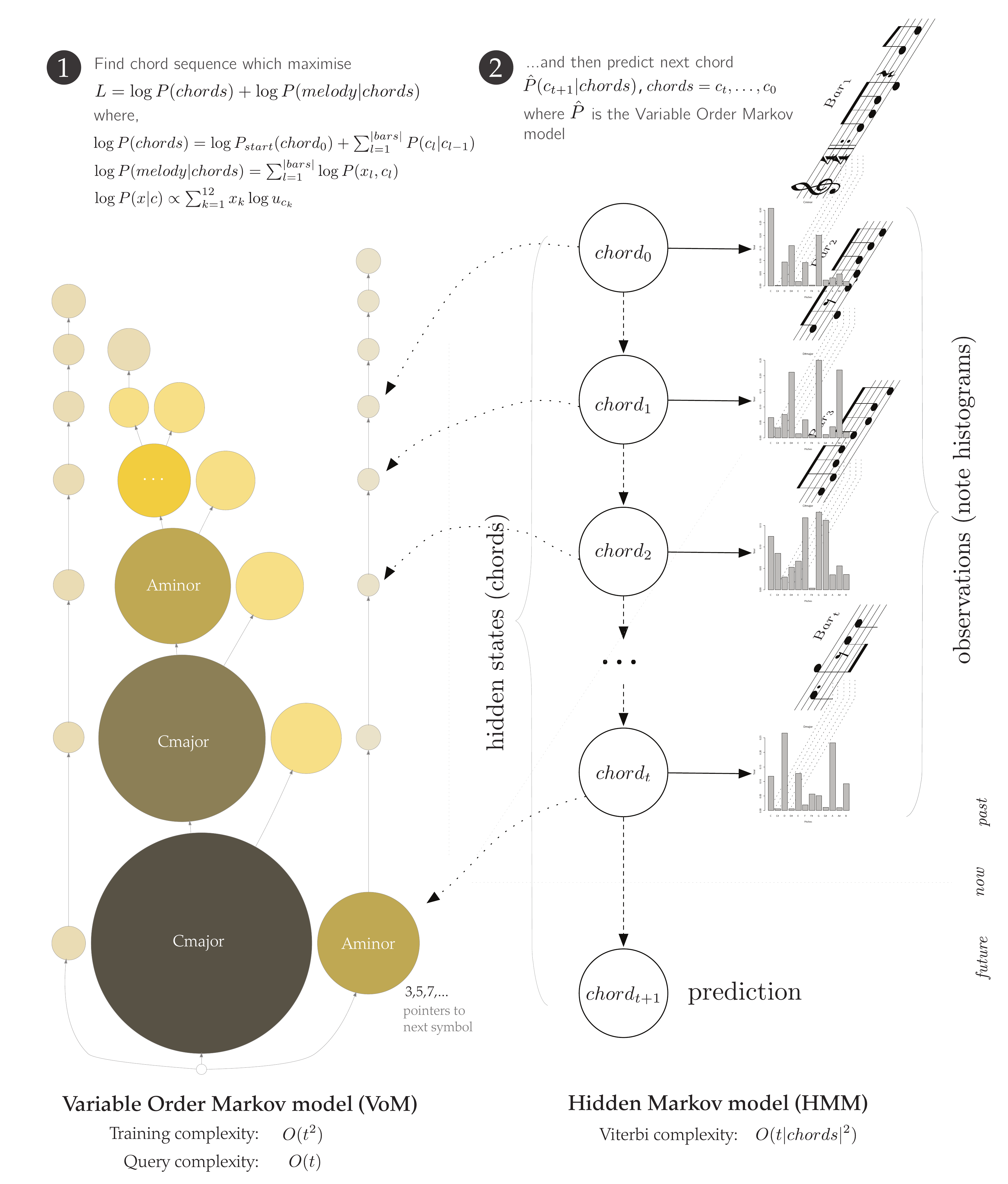}
\caption{The complete system}
\label{system_complete}
\end{center}
\end{figure}

\subsection{Dataset and Training}

So far we have seen a description of the system and it's subsystems. However, one of the fundamental problems of most of machine learning problems is how the training occur. As we said, the idea of this thesis is to simulate the way a new musician learn to play and improvise music. That is, with practise with transcribed music. Thus, our initial aim was to find transcriptions of musical tracks.

Our first choice was to start with MIDI files from well known jazz tracks. Real-Books are books that have been created so as to help jazz musicians to learn and improvise known jazz songs. The process of improvising a jazz song was known in the jazz slang as faking. Our initial data set consisted of 274 MIDI real-book's files with melody and chord track. As we saw, MIDI files contains the raw information of a note and the time that played without any other musical information such as key or chord name. That was a great problem to our approach since we had to find a way
to annotate groups of concurrent played notes to chords. After researching we noticed that it was a problem of itself (\cite{Lee}) that might introduce several errors in our method. For example, if we annotate wrongly several chords then the error would be propagated to the learner and our final system won't be reliable.

For this reason, we abandoned the use of MIDI files and we looked for alternatives.MusicXML are files that contain transcribed music, annotated with musical informations such as chord name, key and time signature, etc. Thus, we found that such data will be sufficient to train our model. Our data set was fetched from the wiki style website wikifonia \footnote{\url{http://www.wikifonia.org}}

Unfortunately, wikifonia did not provide the musicxml files in an archive to download, so we developed a web crawler in python. Using our crawler we fetched 3000 musicxml files. After inspecting them we noted that the files contained several pieces whose annotation violated our assumptions. For example, one of our assumptions was that chord change event can occur on each bar. However, there were pieces that more that two chord changes occurred in one bar. Thus, we had to develop python scripts so as to filter the data set. 

The filtering scripts we developed filtered the 3000 musicxml files as follows. As we already said we didn't want files that contained chord changes more that twice in a bar. What is more, another assumption we had was that no key modulation is allowed. However, we decided that musical pieces with key modulation will not be filtered out but preprocessed and used in the training. The preprocessing we applied was to transform each key modulation in C key. Although that might change the musical feeling of the song we noticed that localy, musical statistical elements remain the same (such as the distribution of pitch classes relative to the key).

Another issue we had to deal with was the amount of different chords and different types of chords. For example, in our data set we might have a Cmajor chord and a Cmaj9, which means the both chord are major chord but the one has also a 9th interval added. However, usually those additional notes in a chord are due to stylistic preferences of the player and not part of the compositions. Thus, we decided to simplify chords to the chord representation we already discussed. That way we keep the information needed for the probabilistic harmonisation and also we decrease the zero-frequency effect. Again, for the chord simplification we developed scripts in python which get as input the musicxml file and export a filtered and processed version of the file.

\begin{figure}[H]
\vspace{15pt}
\begin{center}
\includegraphics[width=0.5\linewidth]{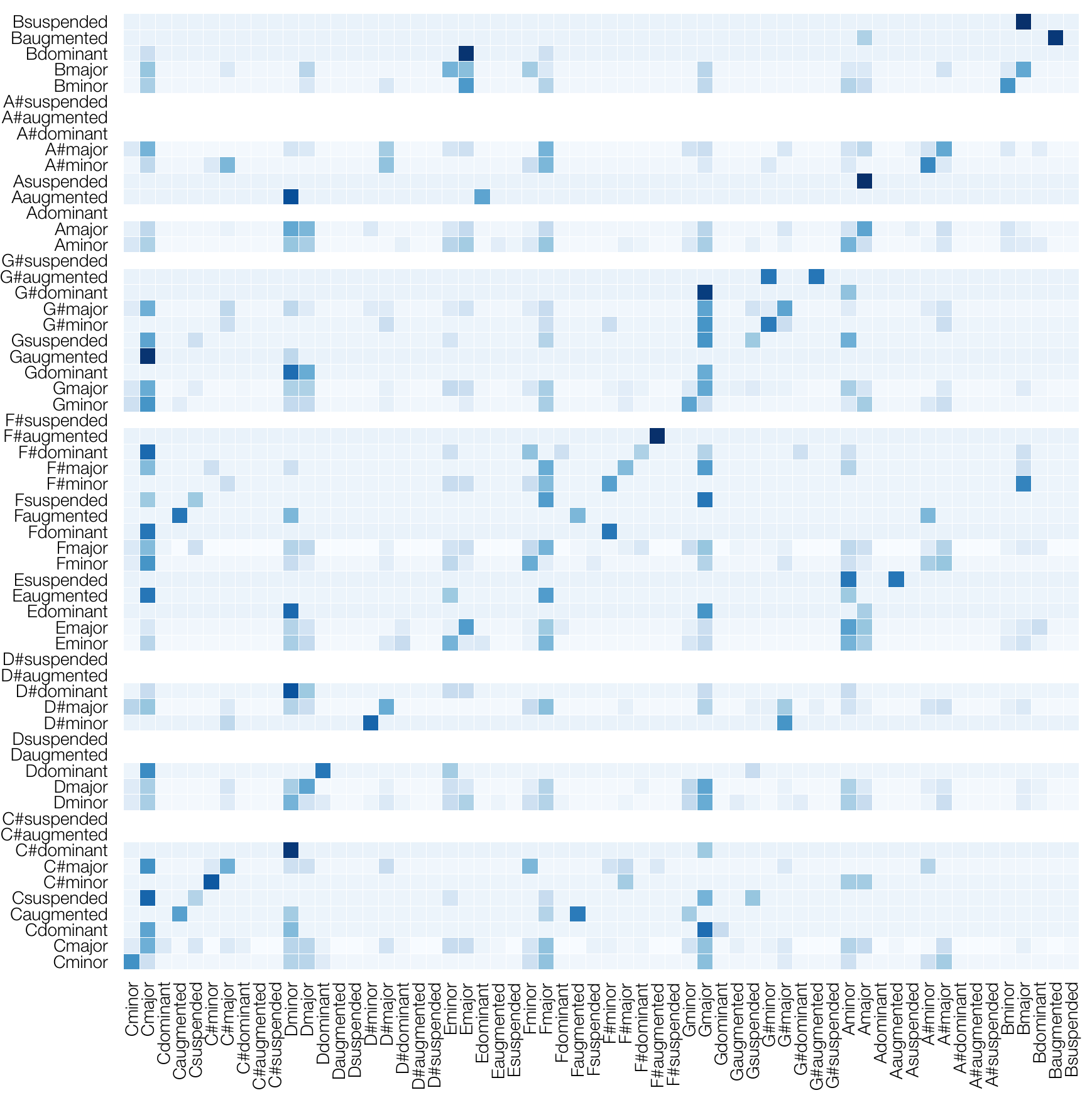}
\end{center}
\caption{chord transition probabilities of HMM heatmatp}
\label{hmm_transition_heatmap}
\end{figure}

Finally, for the training of the system we developed scripts in python for counting chord transition frequencies and initial chord probabilities. The output of those scripts were text files containing transition, initial and emission probabilities. Those files are given to the Hidden Markov Model as input. In figure \ref{hmm_transition_heatmap} we can see the heat map of
transition probabilities as trained from our data.

\subsection{Implementation}

Our system was of great importance to ran as fast as possible. What is more, given the limited time, we wanted a system with which we will be able to develop fast and ideally integrated with existing software of real-time music processing. Given those constraints we decided to develop our system on Max/MSP language.

What attracted us to this language
was the fact that we could route MIDI and audio signals through filters like flowcharts. That way it was easy to prototype ideas and change filtering and routing of signals during performance. What is more, it is extendable in a mean that we can develop plugins (called externals in Max/MSP terminology) using languages such as C, C++ or Java. 

For this project we developed a Max/MSP external. Our whole system was developed using Java and integrated into Max/MSP. After experimenting with several Hidden Markov Model implementations we decided that the best was to develop it from scratch. The reason we decided so was because of the customisability of such choice. What is more, the time it would take us to customise a ready library will be more than developing the library from scratch. For the same reasons we also developed from scratch the Variable Order Markov model. The final external consists of two inputs and one output. The first input takes the notes played by the musician and the second input waits for a signal that the bar has changed. Thus, each time that bar changes, notes gathered from the input are grouped to bars and the processed (e.g. pitch distributions ,etc). What is more, the predict algorithm is executed and the predicted chord is the outputted.

As we said, the whole system was developed as a Max/MSP external with two inputs and one output. Then, Max/MSP allowed us to route midi signals to the system and also route the output to midi devices (such as a midi synthesiser). Further more, Max/MSP enabled us to integrate our system with the commercial live performance software Ableton Live. That way, we could use our system with the state-of-art software in music live performance.

The whole system can be downloaded and used from \url{http://ptigas.com}.

\subsection{Conclusion}

In this chapter we saw the complete description of our system, design decisions we made and why, and description
of our implementation. In the following chapter we will see the different settings we used in our experiments and the results which show the main contribution of this thesis.

\chapter{Testing and evaluation}

\epigraph{\emph{Good art is not what it looks like, but what it does to us.}}{Roy Adzak}

\section{Introduction}

In this chapter we present the experimental settings we used and the evaluation we utilised to measure and compare the performance of our system. Also, we present the results of the experiments and the evaluation methods.

\section{Objective evaluation}

As we have already discussed we had to evaluate our system using objective criteria. Recalling the aim of this project, we had to develop a system that predict chords at real-time. Thus we had to measure the response of the time and the accuracy of the prediction.

\subsection{Time performance}

\begin{figure}[h]
\centering
\includegraphics[width=0.8\linewidth]{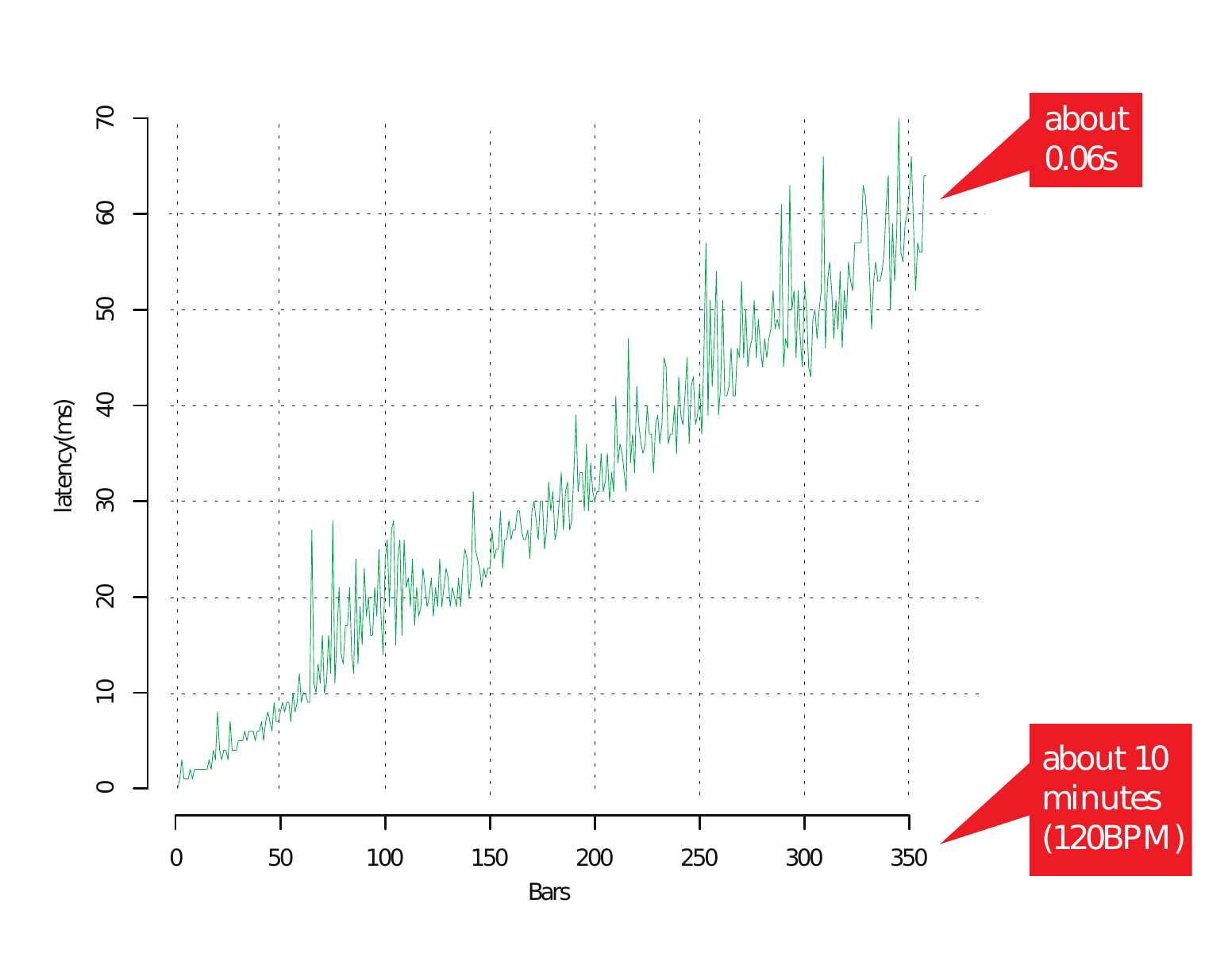}
\caption{System's latency versus session's time}
\label{time_performance}
\end{figure}

We designed our system so as be as fast as possible. The time complexity of both Hidden Markov Model and Variable Order Markov model was small enough to response at real-time. However, complexity by itself is not enough since the actual performance depends on architecture and several other things (bottleneck, scheduling, etc). Thus, we had to measure the actual time it needs the system to respond -- that is predict a new chord. For this, we developed simple Max/MSP externals which allowed us to measure the time between the bar change and the availability of the prediction. We measured the performance of the system during a rehearsal of duration of 10 minutes. That time was enough for an improvisation to complete and also a reasonable amount of time to measure. In figure \ref{time_performance} we can see the response of the system. Please note that response time decrease during playing since both chord sequence inference and prediction algorithm are sensitive to the size of history -- that is the number of bars played.

\subsection{Prediction accuracy}

\subsubsection{Evaluation framework}

The most important objective of this project was to predict future chord using the history of the pitch distributions of the bars. What is more, our data set consisted of chord annotated transcriptions. Thus using the dataset as ground truth we could easily measure the prediction accuracy.

For prediction accuracy measure of a test we used the percentage of chord predictions that were equal to the actual chord in the testing MusicXML. That is,

\[
m(\text{song})=\sum_i^T{L(chord_t, prediction_t)}
\]

 where $T$ is the number of bars of the test, $chord_t$ is the actual chord  of test at bar $t$ and $prediction_t$ is the prediction at bar $t$ using history of pitch probabilities for bars from $0$ to $t-1$.

What is more, we wanted to measure the performance on independent data so as to wee how good our system was in generalising. For this we used a method known as cross validation. This method splits data into $n$ folds ($fold_1,\dots, fold_n$). For $i=1,\dots, n$ we repeat the following. We use $fold_i$ as testing dataset and the rest folds for training. After training the system with data from training fold we measure the performance of each of the tests of the testing data set and then compute the average. Finally we average those averages for the final prediction accuracy of the system. In this thesis we used 10-fold cross validation.

\subsubsection{Experimental settings}

For the evaluation of out system we designed the following experiments.

\noindent
\textbf{1) HMM inference accuracy}\\
In this experiment we aimed to evaluate the inference accuracy of the Hidden Markov Model and Viterbi algorithm. In other words, to measure how good is HMM for finding the most probable chord sequence given a history of bars' pitch distributions. It's easy to see that the prediction power of our system depends highly on how good it inferenced the chord sequence.

\noindent
\textbf{2) HMM and VoM prediction accuracy (60 chords)}\\
In previous chapter we saw that we simplified chord representation so as to have 5 different chord types (minor, major, augmented, diminished, suspended) for each of 12 different root notes (C,C\#,D,D\#,E,F,F\#,G,G\#,A,A\#,B). Thus in total we have $12\times 5=60$ different chords. In this exeriment we used all music pieces that are valid according to our assumptions and whose chords are simplified to those 60 chords.

\noindent
\textbf{3) HMM and VoM prediction accuracy (7 chords)}\\
In the previous experiments we used as data set music pieces that contained chords from 60-chord set. However, recalling music theory, the key and the scale information indicates which notes are most probable to occur. For example, if the key is C then most of the notes in the musical piece will be the notes that consist to C major scale. Those are C,D,E,F,G,A,B. If we use strict music theory then under the fact that we are in C major, only those notes are allowed. Under this constraint, we are allowed to use only 7 chords, called diatonic chords do to their connection to the key of the musical piece.

We decided that this experiment will show how good is our system using the assumption that strict musical theory is applied. For this we filtered again our data set and kept only pieces that used as chords the 7 diatonic chords.

\begin{figure}[h]
 \centering
\begin{tabular}{p{50pt}c}
 Cmajor & \includegraphics[width=0.3\linewidth]{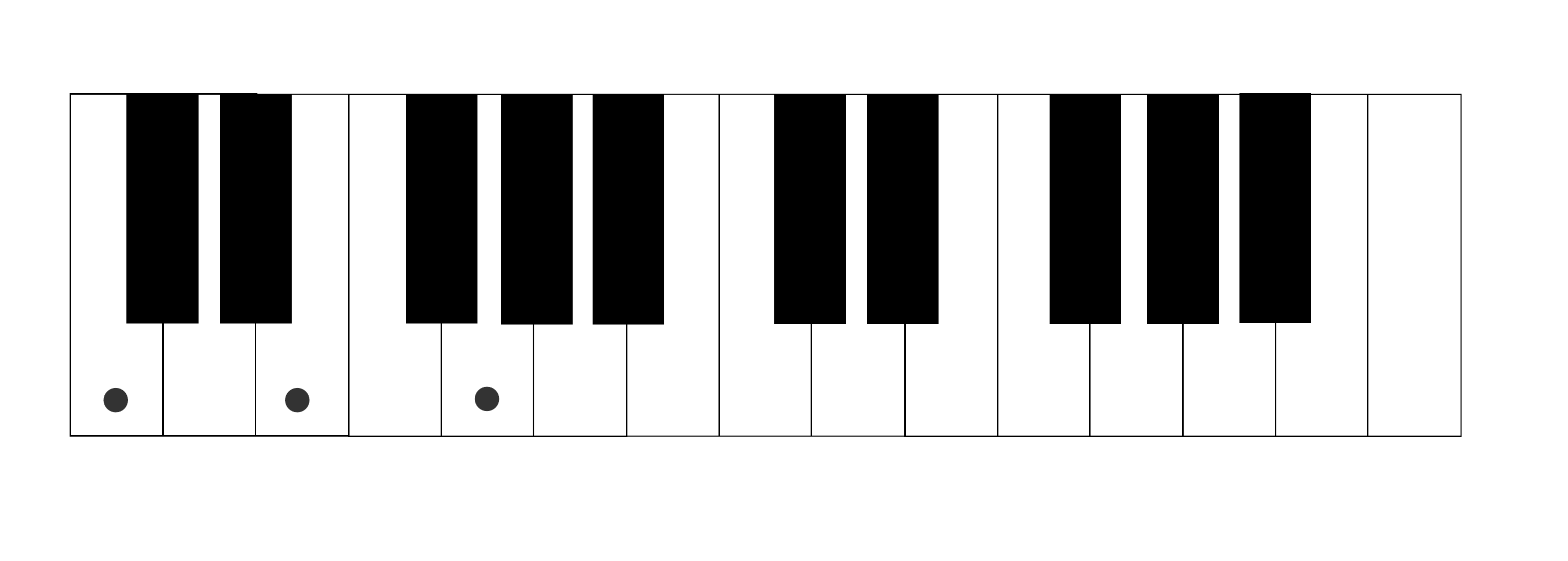} \\
 Dminor & \includegraphics[width=0.3\linewidth]{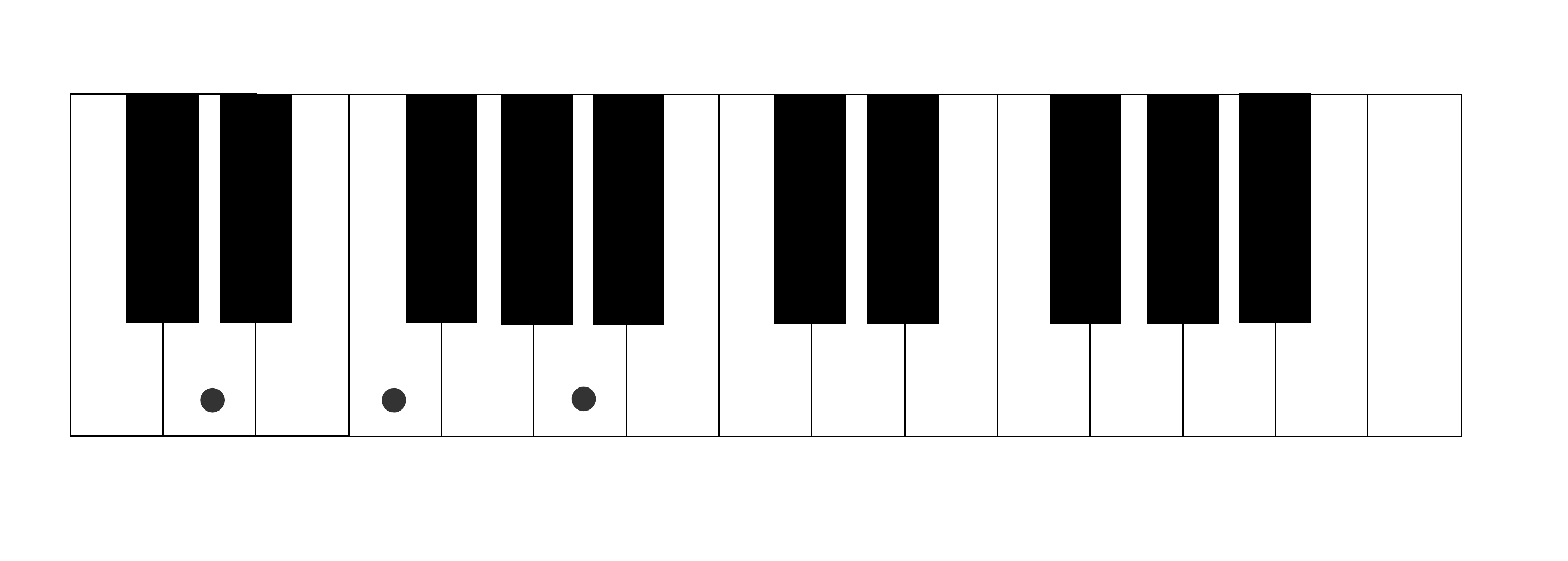} \\
 Eminor & \includegraphics[width=0.3\linewidth]{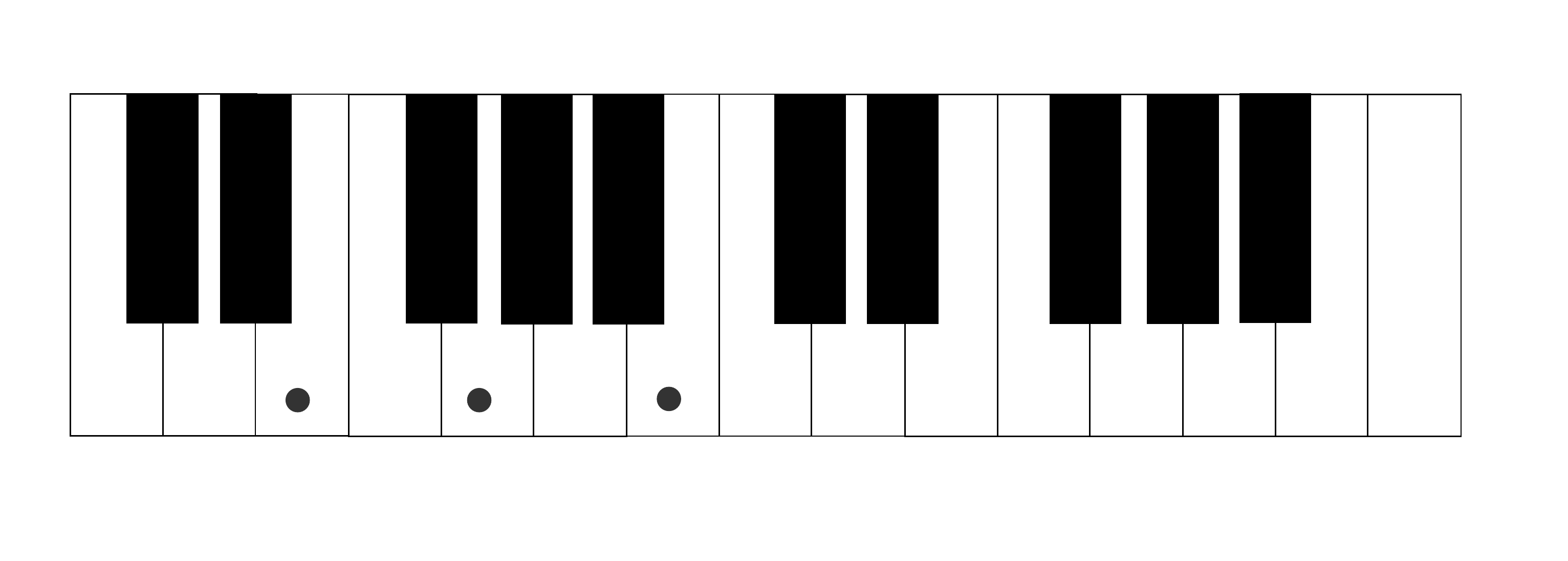} \\
 Fmajor & \includegraphics[width=0.3\linewidth]{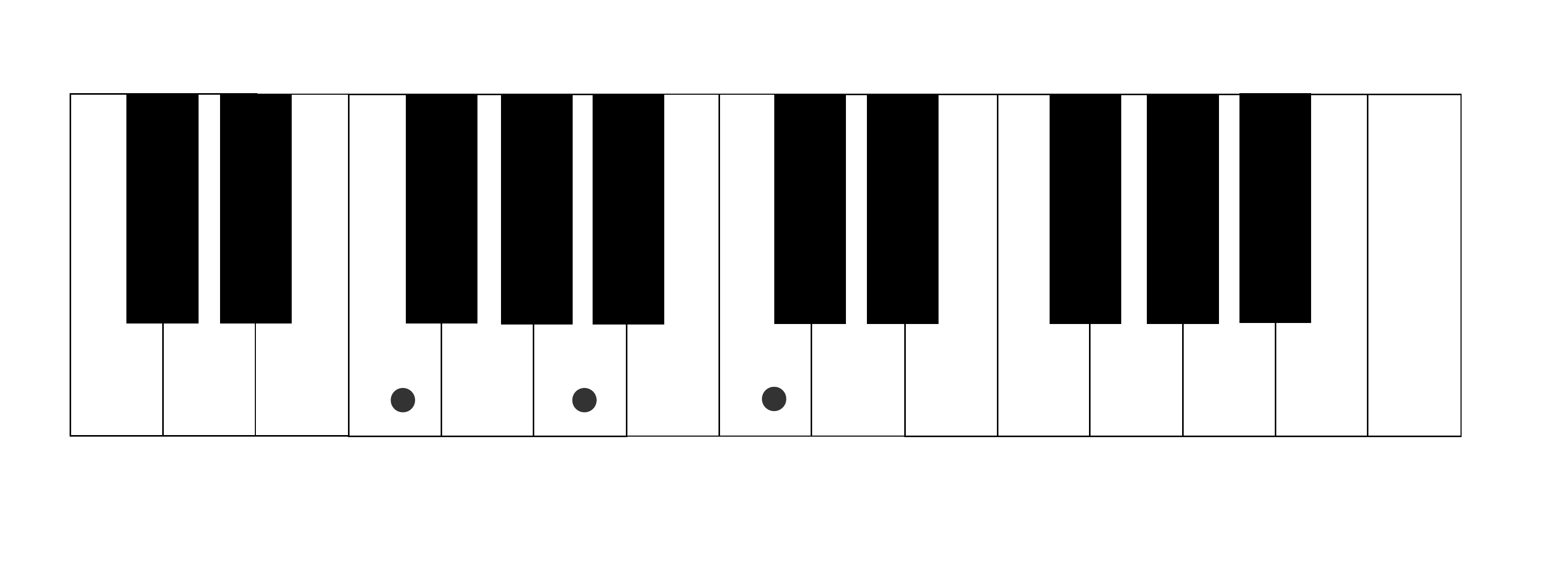} \\
 Gmajor & \includegraphics[width=0.3\linewidth]{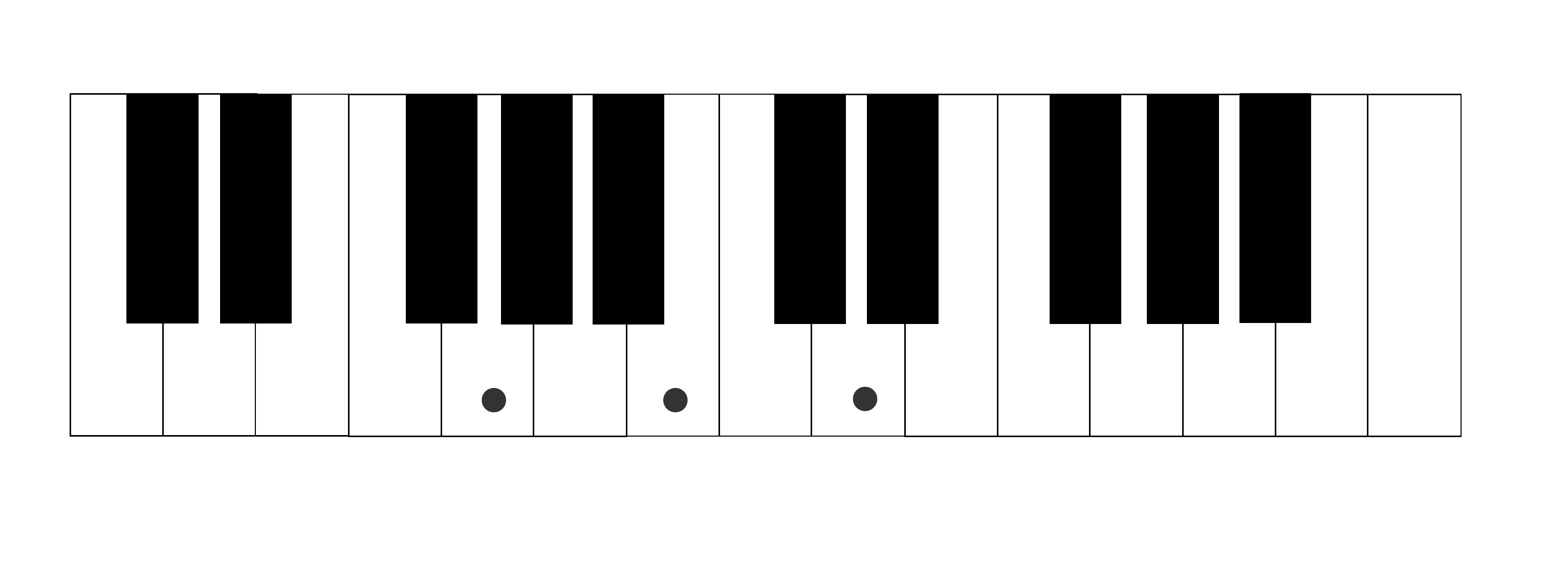} \\
 Aminor & \includegraphics[width=0.3\linewidth]{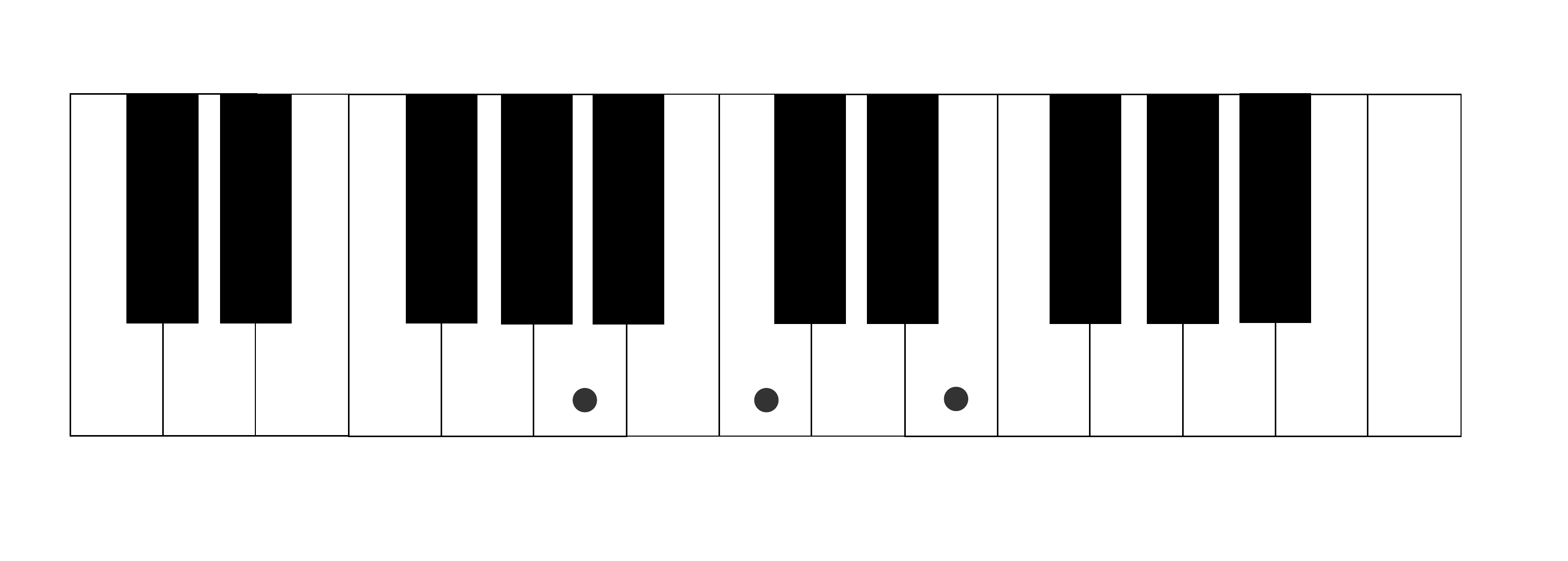} \\
 Bdim & \includegraphics[width=0.3\linewidth]{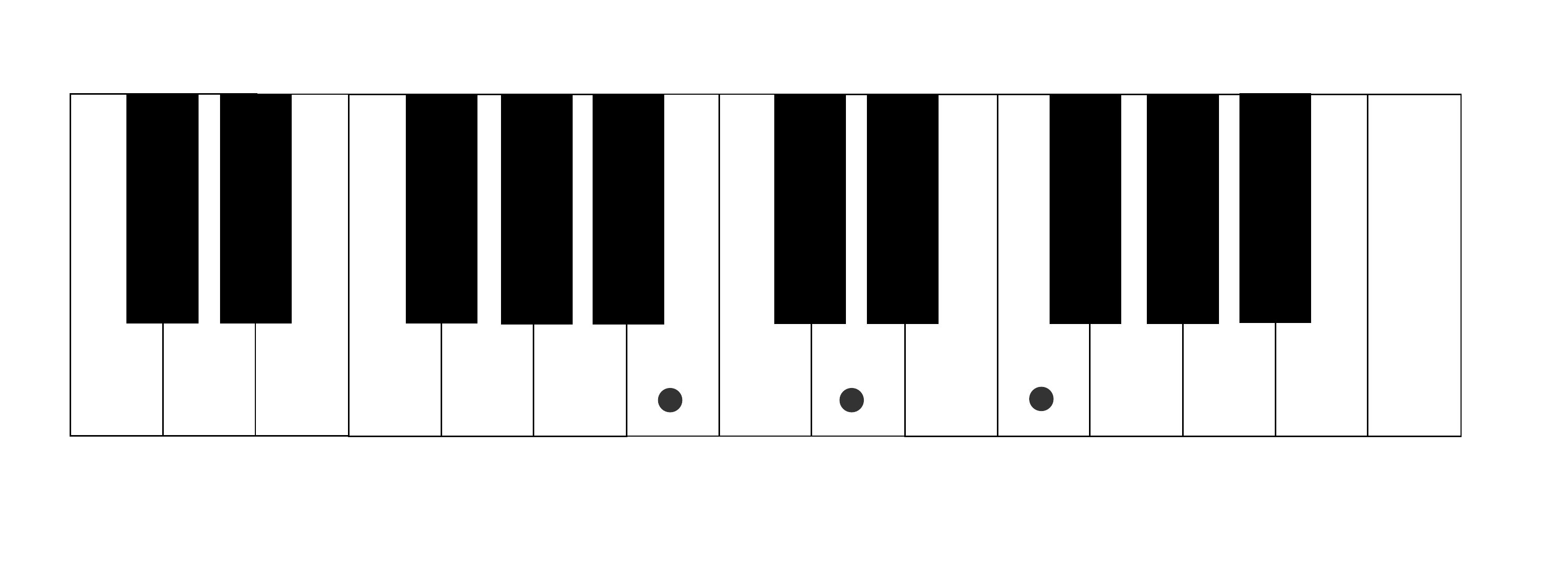}
\end{tabular}
\caption{7 Diatonic chords of key C}
\end{figure}

After the filtering we ended with 235 MusicXML files that had only diatonic chords. This data set is then used for training and testing as discussed.

\noindent
\textbf{4) HMM and VoM half-half}\\
We claim that our system's strength is that it can learn during the rehearsal and improve it's prediction ability. So as to measure this ability of our system we designed the following experiment. 
As we said we measure the prediction performance of our system on a song using the formula $m(\text{song})=\sum_i^T{L(chord_t, prediction_t)}$. In this experiment, instead of measuring the performance on all song, we measure it on the first and second half separately. That is :
\[
 m_1(\text{song})=\sum_{i=0}^{T/2-1}{L(chord_t, prediction_t)}
\]
and
\[
 m_2(\text{song})=\sum_{i=T/2}^T{L(chord_t, prediction_t)}
\]
The results of $m_1$ and $m_2$ are saved into two separate files and processed to calculate p-values and overall prediction accuracy.

\noindent
\textbf{5) HMM using transition matrix for prediction}\\
As we were designing the system, initially we used for prediction the transition probabilities as learned from the corpus. Thus future chords were predicted as follows :
\[
 chord_{t+1}=\arg \max_{c}{P(c|chord_t)}
\]

This experiment actually measure the prediction power of a first-order Markov model trained off-line from corpus. The reason why we wanted to measure the performance of such model was because we wanted to test if the variability of the Markov model increased significantly the performance.

\noindent
\textbf{6) Bayesian Band}\\
Finally, the last experiment was to test the Bayesian Band system. For this, we used the system as it was trained by them. The data set we used for testing were the 235 MusicXML files which contained only diatonic chords.

\subsubsection{Hypotheses testing}
We obtained p-values for the following two null hypotheses.
\\
\noindent
\textbf{H1:} BayesianBand performs better that HMM+VoM (our system).
\begin{center}
and
\end{center}
\textbf{H2:} Our implementation's prediction accuracy is not getting improved during the rehearsal.
\\
\noindent
For H1 and H2 we did an one-sided paired t-test and computed p-values as follows:

\noindent
\textbf{H1:} We kept the accuracy scores as calculated in two separate files. One for our implementation (HMM with 7 chords) and one for Bayesian Band. We decided to use the 7-chord implementation since also their implementation use only 7-chords and thus the comparison is more fair. What is more, we didn't train their system on our data, but we tested it against our dataset. After the calculation of those two score files we used R programming language \footnote{Programming language for statistical calculations and complex graphics plotting} and executed paired t-test (look appendix). The alternative hypothesis was that our system is better than BayesianBand.

\noindent
\textbf{H2:} Using the same workflow we split the prediction accuracy evaluation for the first and second halves of each song. That way we measured if the first half of each song contributed to the improvement of prediction for the next half of each song. The alternative hypothesis was that our system gets improved during rehearsal.

\subsubsection{Results}

For the experiments we developed bash scripts that execute and measure the performance of the system for each experimental settings we discussed. The results obtained can be seen in figure \ref{results}

\begin{figure}[H]
 \centering
 \rowcolors{1}{white}{tableShade}
 \begin{tabular}{p{250pt}|l}
  \textbf{Setting} & \textbf{Performance} \\
  HMM inference & 68.028\%\\
  HMM+VoM (60 chords) & 34.46\%\\
  HMM+Vom (7 chords) & 49.24\%\\
  HMM+Vom (7 chords) first half & 47.8\% \\
  HMM+Vom (7 chords) second half & 50.67\% \\
  HMM with prediction using transition probabilities & 48.76\% \\
  Bayesian Band & 28.03\%
 \end{tabular}
\caption{Results of the experiments}
\label{results}
\end{figure}

\begin{figure}[H]
 \centering
 \rowcolors{1}{white}{tableShade}
 \begin{tabular}{p{250pt}|l}
  \textbf{Hypothesis} & \textbf{ one-sided paired t-test p-value } \\
  Bayesian Band performs better than HMM+VoM & $8.105*10^{-7}$\\
  HMM+VoM prediction accuracy doesn't get improved during rehearsal & 0.0235
 \end{tabular}
\caption{p-values of hypotheses testing}
\label{p-values}
\end{figure}

\section{Subjective evaluation}

Evaluating creative systems only with objective criteria is like ignoring the artistic aspect of such systems. Interactive music systems, as creative systems, aim to output music and music is defined by it's audience. We don't aim to argue on what is music and what is not, however, the musical and artistic aspect of our system cannot be evaluated using other than humans.

For this, we developed questionnaire and asked from the participants to rate the musical result of our system and of Bayesian Band. In detail, we used two compositions and one improvisation. Although, we understand that this is not improvised music, we defined improvise music as music that is unknown to the system ; that is, it hasn't been trained with.

The three composition used are :
\begin{itemize}
 \item Mozart - Andante in C
 \item Beatles - Hey jude
 \item Improvisation in C key.
\end{itemize}

We recorded each system rehearsing with each composition. For this experiment we allowed only 7 diatonic chords because of the nature of BayesianBand system and because we wanted to be as fair as possible.

We wanted to keep the questionnaire as simple as possible so as the participant not to get tired and to decrease any possible bias. For each question we ask the to answer using a simple Likert scale (0 to 4) which is a psychometric scale extensively used in questionnaires. The first question asked was the music skills of the participant. The answers for this question were 0 for no training and 4 for professional musician. After this initial question at each page we have the audio recordings for each system for the same composition. Below each audio clip we ask the participant to rate the quality of the chords. The scale used was 0 for very bad, 1 for bad, 2 for neutral, 3 for good and 4 for very good. We developed three such pages, one for each composition. The order of the recordings is randomised so as to decrease bias introduced by the recognition of some patterns or trends in chords prediction. The following figure presents the raw results of the questionnaire and in next chapter we will analyse the results.

The questionnaire and audio samples can be found in \url{http://ptigas.appspot.com}. Also in the appendix you can find raw results of the questionnaire.

\begin{figure}[H]
\centering
\includegraphics[width=0.5\linewidth]{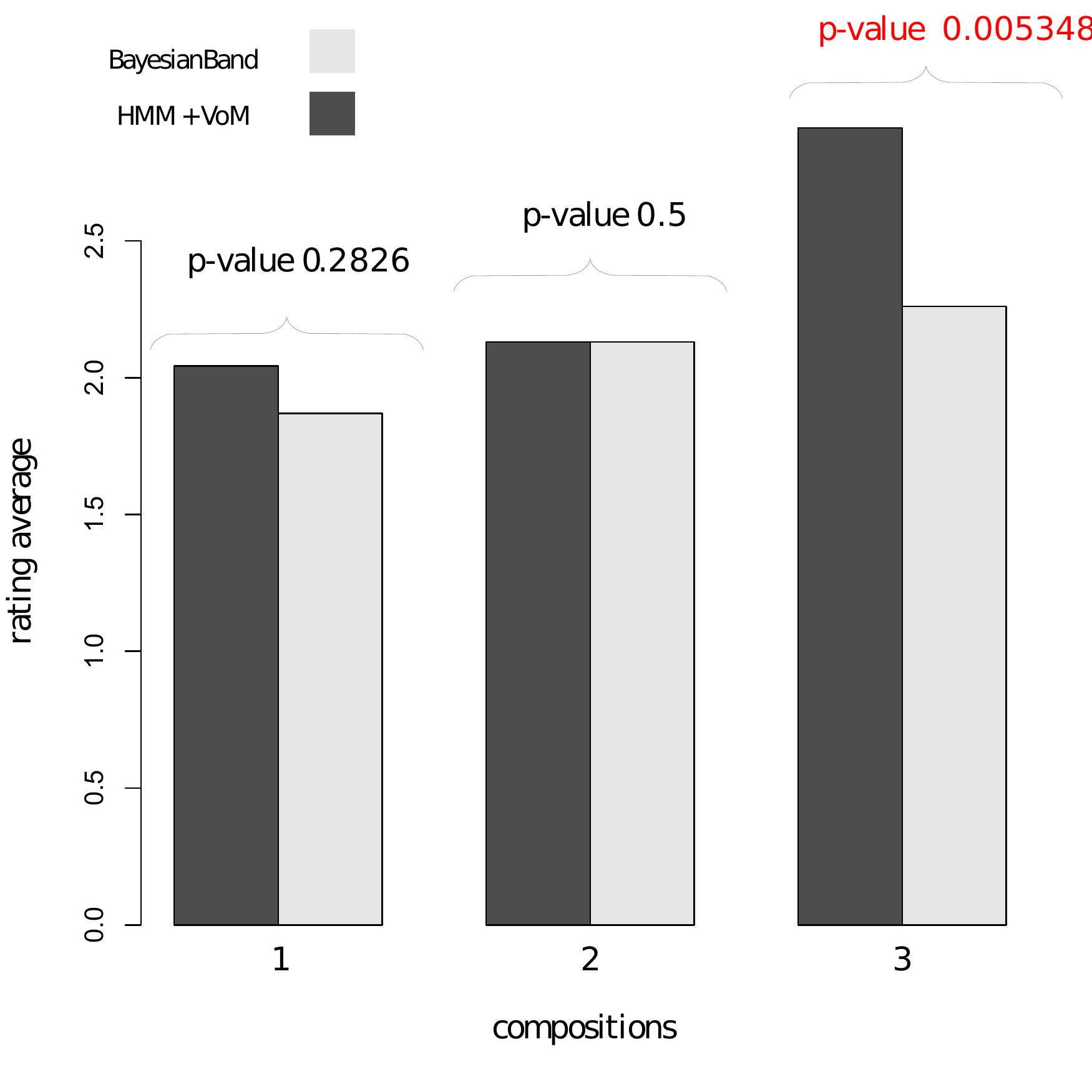}
\vspace{-10pt}
\caption{Average of the questionnaire's ratings per system per composition}
\label{q_averages}
\end{figure}

Again we used hypothesis testing with one-sided paired t-test with alternative hypothesis that our system performs better than BayesianBand.  As we can see, the results in first and second composition doesn't show any significance improvement. Our observation was that in both systems chords showed great variance. For example, in our system chords didn't change frequently and there were cases that it didn't sound very well. In BayesianBand however we noticed the following interesting behaviour. The probabilities of staying in the same chord were very small and thus at every bar it tended to change chord. That sounded more interesting (a participant comment that behaviour "playful") but not always correct. 

Also, keep in mind that each participant had different taste and thus sometimes they preferred our system and some other times BayesianBand. In the third composition the answers showed that participants preferred our system with p-value $0.005348 < 0.05$ which was a significant improvement. Our interpretation of this result was that the melody was polyphonic and thus at each bar the pitch distribution was more informative and thus emission probabilities were calculated better. The result was that our system sounded better and in our subjective opinion chords were more accurate.

\begin{figure}[H]
\vspace{-10pt}
\centering
\includegraphics[width=0.4\linewidth]{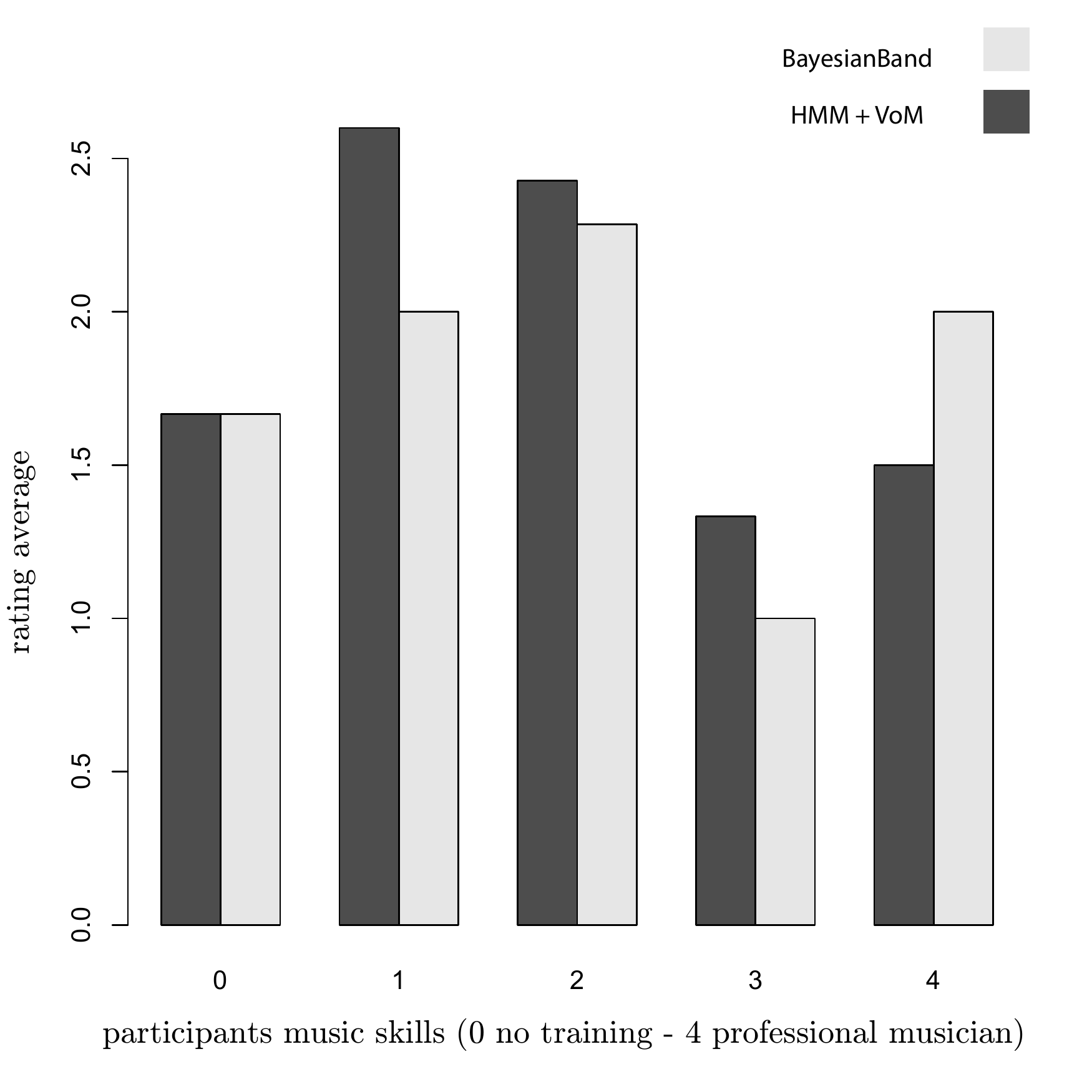}
\caption{ratings per skill for composition 1}
\label{q_c1}
\end{figure}

\begin{figure}[H]
\vspace{-10pt}
\centering
\includegraphics[width=0.4\linewidth]{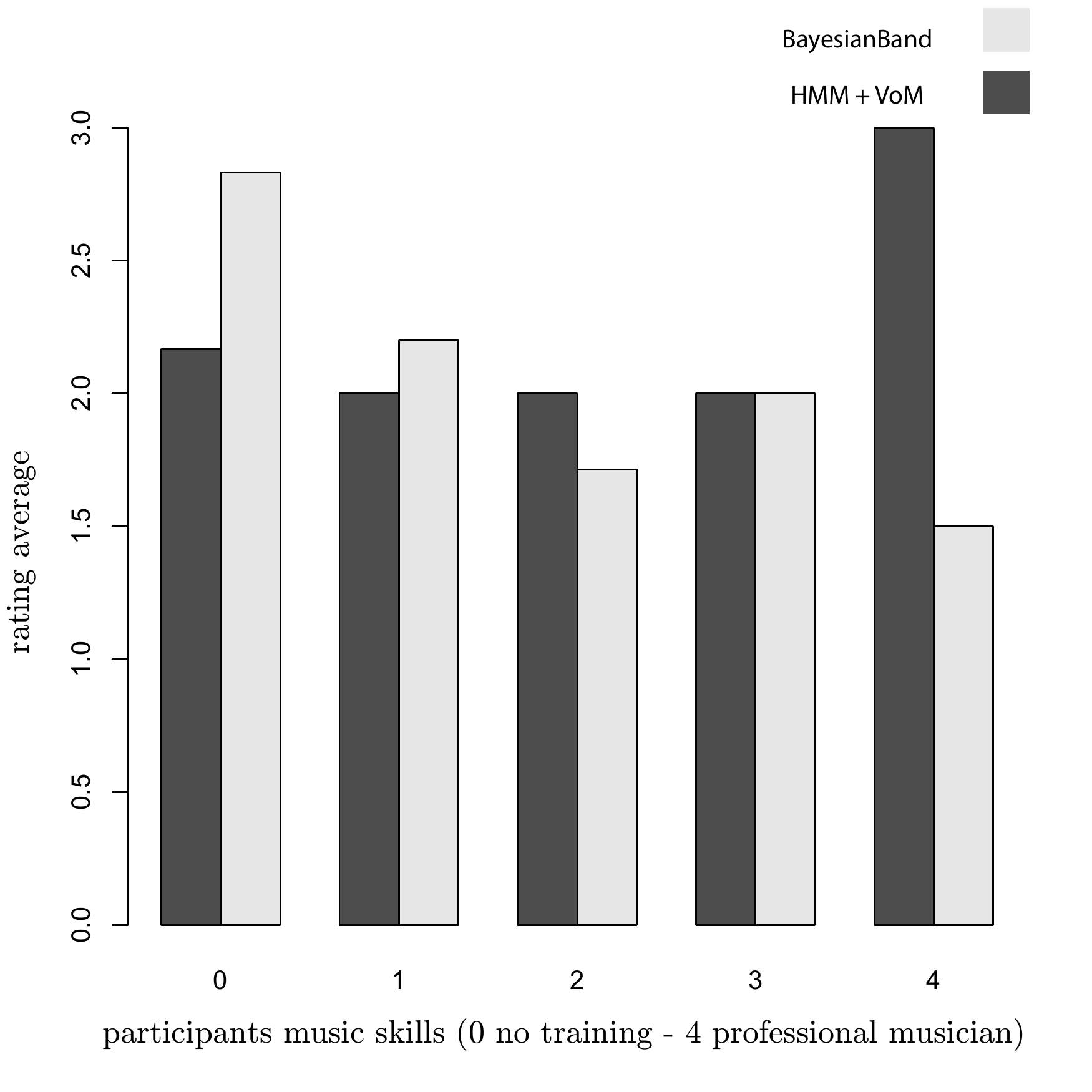}
\caption{ratings per skill for composition 2}
\label{q_c2}
\end{figure}

\begin{figure}[H]
\vspace{-10pt}
\centering
\includegraphics[width=0.4\linewidth]{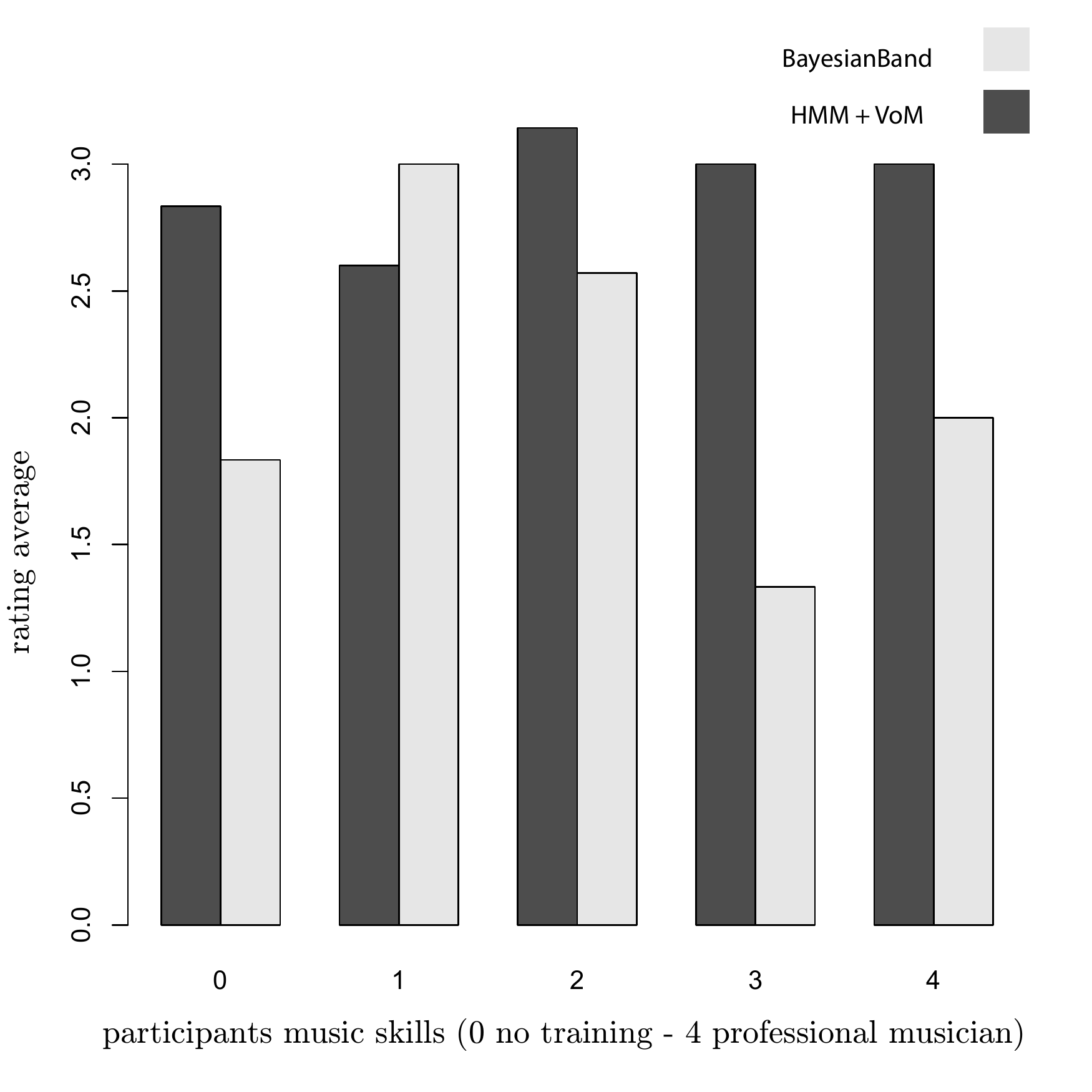}
\caption{ratings per skill for composition 3}
\label{q_c3}
\end{figure}

\chapter{Discussion, conclusions and future work}

\epigraph{\emph{If you develop an ear for sounds that are musical it is like developing an ego. You begin to refuse sounds that are not musical and that way cut yourself off from a good deal of experience.}}{John Cage}

In this chapter we will evaluate to what extend we achieved our goals. Using the experimental results from previous chapter we will critically discuss about the objectives fulfilment and draw a conclusion of this thesis. Finally, the directions we aim to extend our work in the future will be referred.

\section{Aims achievement}

In the Introduction of this thesis we mentioned the aims and objectives of this project. The main goal was to develop a system that is capable of providing accompaniments to an improvising musician during the rehearsal in real-time. Our results indicates that our system predict chords performing better than existing software (BayesianBand) and moreover works in real-time.

As we saw in previous chapter (figure \ref{results}) the prediction accuracy of our system was $34.46\%$ for 60 chords and $49.24\%$ for 7 chords. That indicates that there was an improvement of the performance as we reduced the set of chords to the diatonic chords. One of the reasons why this happened was that less chords implied less choices in chord prediction. Indeed, it's easy to see that the probability of a single chord prediction to be accurate is 1/7 whereas with 60 chords is 1/60. What is more, diatonic chords have the property that they are the most probable chords to occur during a musical piece. Thus, thus, transition probabilities to diatonic chords tend to be greater. When we test our system using 60 chords chances are the we won't be able to predict accurately non-diatonic chords, due to the small probability of transition to such chords. What is more, we noticed that there were diatonic chords that had similar chord-profile with non-diatonic chords. Thus, although we might have a big emission probability for a non-diatonic chord, again, the transition probability bias the preference of our system to a diatonic chord.

The performance of 7-chord system shows that it outperforms BayesianBand system in the objective evaluation which had a prediction accuracy of $28\%$ on the same dataset. Both systems use 7 diatonic chords thus in our opinion the comparison was fair. As we mentioned in previous chapter, we tested the null hypothesis that our system performs worse than BayesianBand in objective evaluation. More formally, the null-hypothesis was that the mean difference of the performances of our system versus BayesianBand's performance is less than zero. One-sided paired t-test showed that with p-value $= 8.105*10^{-7} < 0.05$  we rejected null hypothesis and thus we had a significant improvement \footnote{The alternative hypothesis was that our system is more accurate than BayesianBand}.

One of the greatest achievements of this project was that the system is capable of learning during the performance (online training) and improving it's prediction accuracy. The prediction accuracy of the first halves of the songs is measured $47\%$ and for the second halves $50.67\%$. Thus, our results shows that the first half of each song contains information that our system capture and thus the prediction for the second half of each song gets improved. What is more, the improvement of the performance was significant with p-value $0.0235<0.05$.

Also, our subjective evaluations showed that our system doesn't perform worse than BayesianBand. What is more, in one of the compositions we outperformed BayesianBand significantly with p-value $0.005348 < 0.05$. Those results seem promising although further improvements can be done.

Finally, one important characteristic of our system is that it has to work in real-time. As we saw in previous chapter, our system indeed is capable of providing prediction with latency $<0.06s$ on a 2.0 GHz core 2 duo computer with 2GB RAM. That indicates that our system can indeed work in a real-time context for a simple 10 minute improvisation, although further improvements can be applied. The latency increase during time, which is expected since both complexities of Viterbi algorithm and training of Variable Order Markov model are functions to number of bars played.

\section{Future work}

One of our low priority goals was to develop a system that will actually deal with audio and not only symbolic music (MIDI). In literature there have been many tries to solve this problem however none of which solved this problem efficiently. What is more, transcribing or pitch tracking an audio source is a problem on it self. Thus, it is expected that the implementation of such system would decrease the performance since the error is propagated. Due to limited time, although we developed a system that deals also with audio source, we didn't evaluated and thus didn't described in detail.

In \cite{Stark2009}, Adam Start et al presented the idea of retrieving a chromogram from an audio source in real-time. A great advantage of our design was that we could use those chromograms instead of pitch tracking to retrieve pitch list and then computing again pitch class distributions. What is more, this allowed us also to deal with polyphonic audio source and not only monophonic. Then we route those chromograms to our system the same way we did we pitch distributions, and it's easy to see that they are the same. The author of this thesis performed with such system and found that if the source was noise-free enough then it predicted some pleasing chords. However, we understand that a proper evaluation of such prediction accuracy is needed. Thus, one of our future goals is to improve the system and evaluate it again with both subjective and objective evaluation methods.

Another issue that we aim to work on in the future is to experiment with a more complicated model, called Variable Order Hidden Markov model. Since it is relative new model in the field of machine learning, it hasn't been applied extensively, although when it got applied it's results where promising. To our surprise, a recent research from Georgia tech used this model for predicting Table keystrokes in Indian music \cite{Chordia}. Our aim is to extend and compare to the current implementation.

Finally, we noticed that our current implementation was very memory consuming. That was caused by the nature of Variable Order Markov model and primary by the fact that we didn't make any optimisation to reduce memory usage. However, it's easy to see that our implementation is very similar to a suffix tree. What is more, there exist data structure that implements a Variable Order Markov model that is called probabilistic suffix tree and one of our aims is to use a suffix array data structure which is known for it's efficient memory complexity \cite{Ko}.

\section{Conclusions}

Our main and most important goal was to implement a system that predict accompaniments in a jam-session in real time. For this we developed a system that achieves that goal to the extend that a summer project allowed us. Our results indicates that indeed our implementation can run in real time and predict chords with very good prediction accuracy. What is more, in the objective evaluation it outperforms current system called Bayesian Band. Also, our implementation shows that it is capable of learning underlying structure of the improvisation and utilise this knowledge for improving chord prediction. Finally, our design allows us to extend further the system so as to use polyphonic audio noise-free source (e.g. dry signal of a guitar).

In the previous chapters we also reviewed some of the state-of-art interactive music systems (Chapter 2). What is more, in chapter 3 we reviewed the graphical models and analysed the Hidden Markov Model and Variable Order Markov model that we used in this thesis. In chapter 4 we described our implementation along with the decisions we made during this project. Additionally, in chapter 5 we analysed the evaluation methods we used in this project and we presented our results. Finally, in chapter 6 we critically evaluated our results, we gave some directions for further work and we concluded to which extend our aims and goals got achieved.

Last but not least, this thesis fulfilled the ultimate goal of the author which was to combine computer science, mathematics and music. As a byproduct of this thesis was the better understanding of how humans percept music and act during improvisation. As a last comment on this work, we would like to use Johny Cage words\\
\emph{"If you develop an ear for sounds that are musical it is like developing an ego. You begin to refuse sounds that are not musical and that way cut yourself off from a good deal of experience."}\\
Interpreting this in a machine learning context, always such system will reflect the aesthetics of the author and data used for training. However, during this thesis we gained the confidence to say that it's not far the future where humans and machines will be able to collaborate for creating art, although still there are elements of human perception that must be incorporated.

\appendix

\appendix

\chapter{Questionnaire raw results}

\begin{figure}[H]
\small
\centering
\rowcolors{1}{white}{tableShade}
\begin{tabular}{ccccccc}
\hline
\textbf{Music skills} & \textbf{C1 BB} & \textbf{C1 HMM+VoM} & \textbf{C2 BB} & \textbf{C2 HMM+VoM} & \textbf{C3 BB} & \textbf{C3 HMM+VoM} \\
\hline 
1&2&4&2&1&3&1\\
3&0&3&2&3&1&4\\
2&4&2&1&2&2&3\\
0&1&2&4&3&3&4\\
0&3&3&4&2&2&2\\
4&3&2&1&3&2&4\\
0&3&1&3&1&0&3\\
2&2&3&1&2&3&4\\
2&3&3&3&2&2&3\\
1&1&4&3&3&3&4\\
0&1&1&3&2&2&3\\
0&1&2&2&3&1&3\\
0&1&1&1&2&3&2\\
2&3&2&3&3&3&3\\
2&2&2&3&3&3&3\\
1&3&2&2&2&3&3\\
1&2&2&3&3&3&3\\
2&1&3&1&0&3&3\\
2&1&2&0&2&2&3\\
3&2&1&3&2&2&3\\
4&1&1&2&3&2&2\\
1&2&1&1&1&3&2\\
3&1&0&1&1&1&2\\
\hline
\end{tabular}
\caption{Raw questionnaire results. $C_i$ stands for composition $i$, BB for BayesianBand and HMM+VoM for our system.}
\end{figure}

\phantomsection \addcontentsline{toc}{chapter}{Biliography}

\phantom{
\cite{Eck1920}
}

\bibliographystyle{abbrv}
\bibliography{bibliography}

\cleardoublepage

\end{document}